\documentclass[twocolumn]{article}
\usepackage[margin=0.8in]{geometry}
\usepackage[T1]{fontenc}
\usepackage[utf8]{inputenc}
\usepackage{lmodern}
\usepackage[english]{babel}
\usepackage{graphicx}
\usepackage{amssymb}
\usepackage{overpic}

\usepackage{hyperref}
\hypersetup{
    colorlinks=true,
    linkcolor=black,
    citecolor=black,
    filecolor=black,
    urlcolor=black,
    linkbordercolor = {1 1 1},
}

\usepackage{algpseudocode}
\usepackage{algorithm}

\usepackage[square,numbers]{natbib}

\title{The rich still get richer: Empirical comparison of preferential attachment via linking statistics in Bitcoin and Ethereum}
\author{D\'aniel Kondor\,$^{1,\ast}$, Nikola Bulatovic\,$^{2}$, J\'ozsef St\'eger\,$^{2}$, Istv\'an Csabai\,$^{2}$, G\'abor Vattay\,$^{2}$\\
		\normalsize{$^{1}$Singapore-MIT Alliance for Research and Technology, Singapore}\\
		\normalsize{$^{2}$Department of Physics of Complex Systems, E\"otv\"os Lor\'and University, Budapest, Hungary}\\
		\normalsize{$^\ast$ E-mail: \texttt{dkondor@mit.edu}}}
\date{\today}

\begin{document}

	\maketitle

	\section*{Abstract}
		
		Bitcoin and Ethereum transactions present one of the largest real-world complex networks that are publicly available for study, including a detailed picture of their time evolution. As such, they have received a considerable amount of attention from the network science community, beside analysis from an economic or cryptography perspective. Among these studies, in an analysis on the early instance of the Bitcoin network, we have shown the clear presence of the preferential attachment, or ``rich-get-richer'' phenomenon. Now, we revisit this question, using a recent version of the Bitcoin network that has grown almost 100-fold since our original analysis. Furthermore, we additionally carry out a comparison with Ethereum, the second most important cryptocurrency. Our results show that preferential attachment continues to be a key factor in the evolution of both the Bitcoin and Ethereum transactoin networks. To facilitate further analysis, we publish a recent version of both transaction networks, and an efficient software implementation that is able to evaluate linking statistics necessary for learn about preferential attachment on networks with several hundred million edges.

\section{Introduction}

Cryptocurrencies have presented a disruptive change for both economics and computer science. Over the past years, interest in cryptocurrencies resulted in a huge amount of money invested in them~\cite{Baur2018} and a growing amount of research carried out on diverse application possibilities of the underlying technologies, e.g.~blockchain and decentralized trust~\cite{Zheng2016, Bonneau2015,Seres2020}. At the same time, cryptocurrencies provide a unique opportunity as financial systems where the whole list of transactions is exposed, making possible to study the dynamic interactions taking place in them~\cite{Kondor2014b,Phetsouvanh2019,Oggier2020,Wu2020}. This way, cryptocurrencies present a unique perspective by providing the complete history on how novel, alternative financial systems evolve from their inception~\cite{Seebacher2018,Dixon2019}. Furthermore, the appearance of cryptocurrencies has helped research connecting network information with economical analysis to gain momentum due to the availability of high volume data~\cite{Kondor2014a,Gurcan2018,Akcora2018,Kurbucz2019}. 

Considering the list of transactions as an evolving network, cryptocurrencies present one of the largest real-world networks that can be analyzed by the scientific community, with several hundred million total edges. This can be of interest in itself, as it allows to test theories about evolving and time-varying networks on large scales with better statistical confidence. While there is significant interest in how cryptocurrencies work from a network science perspective~\cite{Liang2018,Motamed2019,Wu2020}, we still do not have a comprehensive understanding of which are the relevant processes that shape their network structure. 

In the current study, we evaluate key network characteristics on the transaction networks of Bitcoin and Ethereum, the two most popular cryptocurrencies. We specifically look at network evolution and the dynamics of how nodes gain new transaction partners and gain or lose balance. We build on our previous work~\cite{Kondor2014b} that focused only on the initial phase of Bitcoin and found that preferential attachment drives the evolution of the transaction network and concentration of wealth. Considering the scale of Bitcoin and the many factors influencing transaction dynamics, it is remarkable how well power-law degree distributions and preferential attachment describe its evolution. In the current work, we extend our previous analysis to a significantly longer period of trading with multiple up- and downturns in the market for both Bitcoin and Ethereum; in the case of Bitcoin, this means an almost 100-fold growth in total network size. This allows us to test if the main transaction dynamics found previously stay significant during a timeframe when cryptocurrencies gained several orders of magnitude in total investment and became a main market component instead of just a niche. We show that a process of preferential attachment continues to be determinant for both cryptocurrencies and is robust with regard of the time period analyzed and the method used to reconstruct the transaction network.

We download and process the transaction history of both Bitcoin and Ethereum and reconstruct the temporally evolving transaction network. Since the main components of the network are the transactions which are instantaneous events, there are multiple possible choices for defining a network among the addresses. We show that the activity of addresses is characterized by fat-tailed distributions both in terms of temporal extent, number of transactions they participate in and addresses they come in contact with. Most addresses are short lived according to the practice of users of frequently generating new addresses to obtain increased privacy, while some addresses participate in an especially large number of transactions over an extended time range, giving rise to power-law degree distributions in the aggregated network.

We perform a more in-depth analysis of transaction dynamics, testing how preferential attachment can explain the broad degree distributions seen in the aggregated transaction networks. We evaluate statistics of new edge formation using the rank function methodology developed in our previous work~\cite{Kondor2014b} using different levels of temporal aggregation, testing also the robustness of results. During our analysis, we perform an in-depth comparison among Bitcoin and Ethereum, focusing on comparing the transaction dynamics of regular addresses in the two systems and between addresses and smart contracts in Ethereum.

\section{Methods}

\subsection{Data collection}

\subsubsection{Bitcoin}

We adapted the Bitcoin Core client program (version 0.19) by adding functionality to write out data about transactions and blocks in a CSV format\footnote{Source code of our modified client is available at~\url{https://github.com/dkondor/bitcoin/tree/0.19}}. We used this client to download and extract the blockchain on February 7, 2020. Our data includes 616,345 blocks with 500,663,153 transactions among 609,963,452 unique addresses in total.

We construct a network among addresses by creating a directed edge between each input and output address for each transaction, excluding self-edges. The resulting network has 3,648,627,182 unique edges, that appear 4,834,306,446 times in total. Note that in Bitcoin, a transaction can have multiple input and output addresses and thus can result in the addition of multiple edges~\cite{Phetsouvanh2019}. Also, transaction inputs must always include the full amount received by a previous transaction output; when spending less than this amount, the remainder (or ``change'') is directed to one of the addresses of the spending user in a separate transaction output. This results in a large number of self-edges in practice.

\subsubsection{Ethereum}

We use the OpenEthereum client to synchronize with the blockchain and then use the Ethereum-ETL client to output a the transaction history in CSV format. We extracted data on February 2, 2020; this includes the first 9.4 million blocks in the chain, with a total of 628,810,973 transactions among 68,429,208 unique addresses. Ethereum transactions are one-to-one: each transaction has only one input and output address and thus can be directly mapped to a directed edge in a network among addresses. Contrary to Bitcoin, in Ethereum, the balance of an address is recorded as an intrinsic property in the system; this way, spending is possible in any denomination, and does not require the ``change'' mechanism used in Bitcoin.

\subsection{Edge lifetime}

In the usual picture of growing complex networks, edges are typically considered static entities that represent existing connection which can be gained or lost over time. For transactions in cryptocurrencies, this picture is not accurate: since transactions are instantaneous events, the presence of an edge in our network indicates that at least on transaction took place between two addresses over the lifetime of the network. Given the timescales in our analysis, edges that correspond to transactions that happened a long time ago lose their relevance (e.g.~if a user abandons using a certain address, as is often the case). To account for this, we can use an alternate network definition, where edges have a finite ``lifetime'': they are created when a transaction happens between two addresses, and are removed if a certain time passes without repeated transactions between the same pair of addresses. Removal of an edge also decreases the network degree of the associated nodes. This means that activity is gradually ``forgotten'', at least for the purpose of our analysis.

In this case, the indegree of a node naturally represents the number of distinct transaction partners it had in a recent time interval. We can choose this time interval to correspond to a presumption of ``memory'' in the dynamics between addresses. In practice, we created networks where the lifetime of edges was limited to one day and 30 days beside the fully time-aggregated network.

\subsection{Preferential attachment}

Preferential attachment is a model of network evolution originally suggested by Barab\'asi and Albert~\cite{bamodel}, based on the models studied originally in different contexts by Yule and Simon~\cite{Yule1925,Simon1955}. The original model predicts a power-law degree distribution with an exponent of $\gamma = 2$; it was later generalized to yield networks with power-law degree distributions of arbitrary exponents~\cite{Dorogovtsev2000}. Preferential attachment was observed either directly or indirectly in many real-world complex networks in the past decades~\cite{barabasipa2,paonline,Perc2014}, including an early phase of Bitcoin~\cite{Kondor2014b}.

In this paper, we focus on a model of nonlinear preferential attachment~\cite{Krapivsky2000}, represented in our case by the simple rule that the probability of a new link connecting to a target node with indegree $k$ is proportional to $k^\alpha$. Note that we do not restrict this process to links from new nodes, as we expect a significant amount of links to be created between already existing nodes, a departure of the original Barab\'asi-Albert model~\cite{bamodel}. Also, the choice is made among \emph{existing} nodes, thus the total probability of connecting to \emph{any} node with indegree $k$ is

\begin{equation}
	\Pi(k) \sim n (k) k^\alpha \label{eq:model}
\end{equation}

where $n(k)$ is the number of nodes with indegree $k$ in the network (i.e.~the empirical degree distribution). In an evolving network, the degree distribution will change over time, making it difficult to compare probabilities of events that occur at different times with different network configurations. We overcome this problem by calculating the \emph{transformed rank} of the target indegree for each linking event:

\begin{equation}
	R \equiv \frac{\sum_{k = 0}^{k_\mathrm{target}} n(k) k^\alpha}{\sum_{k = 0}^{k_\mathrm{max}} n(k) k^\alpha}
	\label{eq:R}
\end{equation}

where $k_\mathrm{target}$ is the indegree of the node receiving the new link. If our assumption about the preferential attachment process and the $\alpha$ exponent holds true, then empirical $R$ values calculated for a set of linking events will be distributed in a uniform way over the $[0,1]$ interval~\cite{Kondor2014b}. Since the $R$ transformed rank values are normalized this way, values from different time points (and thus different stages of the evolving network) can be analyzed together. Furthermore, by limiting the set of events considered to smaller time intervals, the role of the preferential attachment process in network evolution at different times can be easily compared.

In practice, we can calculate transformed ranks for any value of the $\alpha$ exponent. In this article, we compare several $\alpha$ values and identify the one that best fits a uniform distribution. Note that a hypothesis of no preferential attachment (i.e.~a case where network degree does not affect the probability of attracting new transaction partners) can be readily represented in this framework by $\alpha = 0$.

Evaluating the statistics of preferential attachment requires calculating the $R$ value in Eq.~(\ref{eq:R}) for each ``event'', i.e.~possible multiple times for each transaction, based on the actual degree distribution in the network. Since the number of transactions is in the order of hundreds of millions for both networks, a direct summation over the degree distribution (that has a runtime complexity of $O(N)$ for a network of $N$ nodes) is not feasible. However, using a properly augmented binary search tree as the data structure to store the degree distribution along with partial sums of $k^\alpha$, we are able to perform the calculation of $R$ values in $O(\log N)$ time complexity, making it possible to evaluate the distribution of $R$ values over hundreds of millions of events. We describe the necessary tools used for this purpose in the Supplementary Material, while we publish the source code of an efficient augmented binary search tree implementation used for this purpose online~\cite{orbtree,patest_new}.

\section{Results}

\subsection{Network growth and structure}

Both Bitcoin and Ethereum has experienced a great amount of growth over their lifetime, including multiple ``peaks'', where a sudden surge of interest resulted in large upticks of both exchange price and network activity. Since early 2018 when cryptocurrencies gained an unprecedented global attention, daily activity for both Bitcoin and Ethereum has had an approximately constant rate however, in contrast to previous periods of growth. This could be the consequence of getting close to the technical limits of transaction volume that the networks are able to handle\footnote{Both Bitcoin and Ethereum have hard limits on the amount of data, and thus the number of transactions that can be included in blocks (Bitcoin directly limits the block size, while Ethereum limits the maximum gas amount to be used in blocks). Approaching this limit will result in transaction fees increasing (since miners will prefer to include transactions with more fees). This functions as a natural feedback loop that discourages creating too many transactions and thus limits the network activity.}. Also, since the beginning of 2018, the total capitalization of cryptocurrencies (for simplicity, defined as the total value of coins in circulation based on the current exchange rate) have approached that of publicly traded stocks with the highest capitalization; this could limit further speculative investment in them.

\begin{figure*}
	\includegraphics{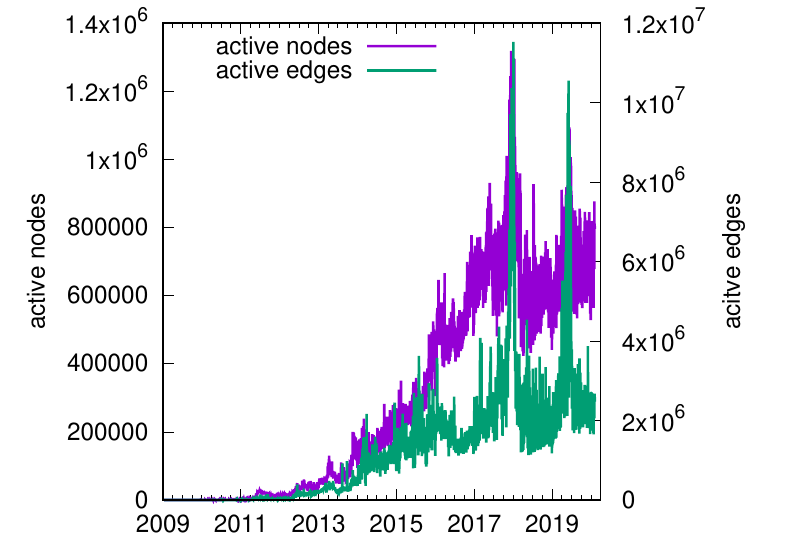}
	\includegraphics{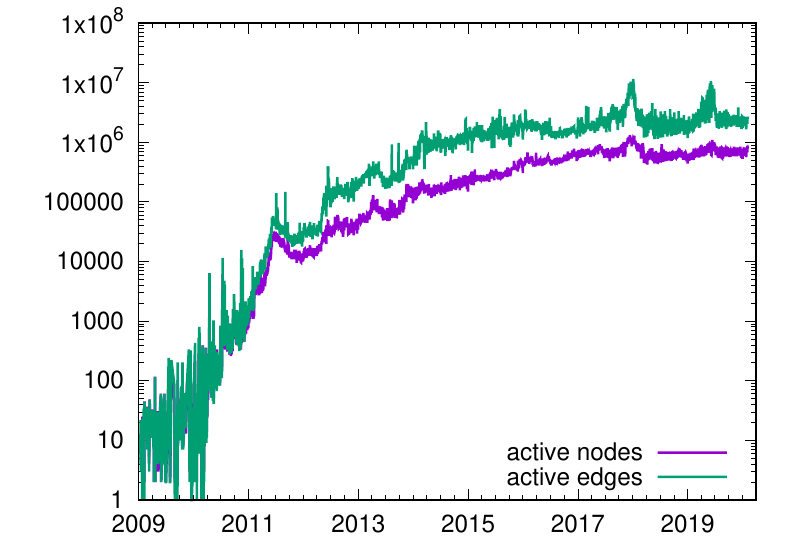}
	\caption{Timeline of activity in the Bitcoin network, measured by the number of nodes (addresses) and edges active each day on a linear (left) and logarithmic (right) scale. We see that the activity in Bitcoin experienced a steady growth over several years after an initial surge of interest in 2011. In the recent years, growths has tapered off, with activity stabilizing around a few million edges per day.}
	\label{bitcoin_activity}
\end{figure*}

\begin{figure*}
	\includegraphics{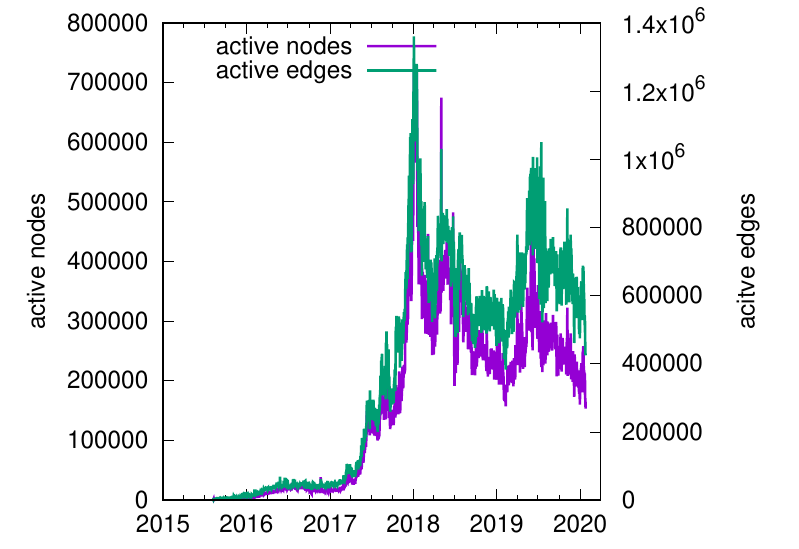}
	\includegraphics{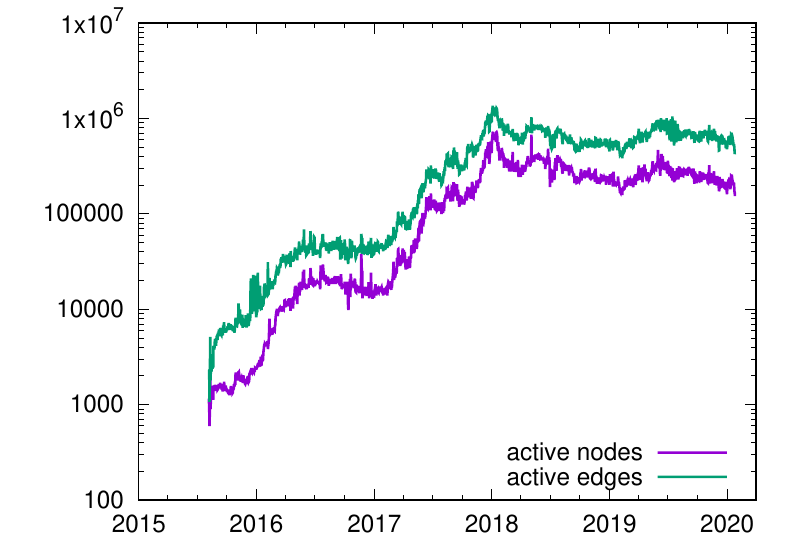}
	\caption{Timeline of activity in the Ethereum network, measured by the number of nodes (addresses) and edges active each day on a linear (left) and logarithmic (right) scale. Growth of activity here is characterized by two distinct phases: an approximately exponential growth phase in the first $2.5$ years, followed by an approximately constant level of activity in the past years.}
	\label{eth_activity}
\end{figure*}

\begin{figure*}
	\includegraphics{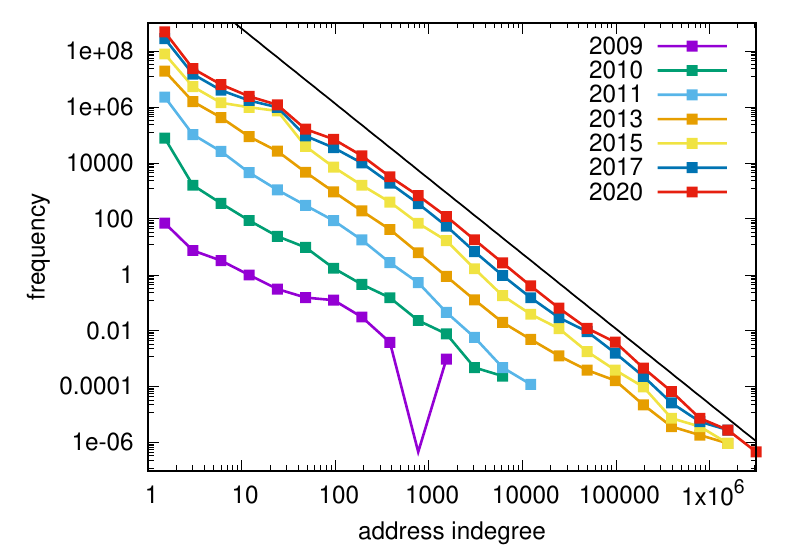}
	\includegraphics{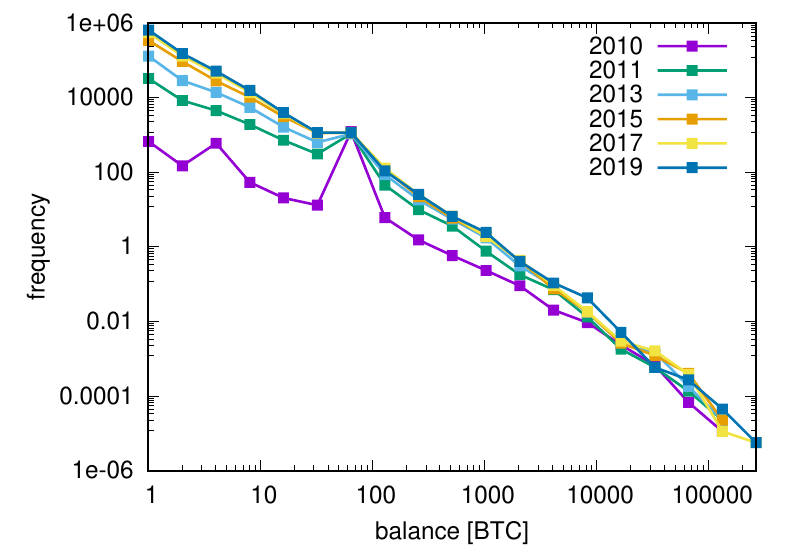}
	\caption{Distribution of network indegrees (left) and address balances (right) for Bitcoin. Indegrees are determined by the total number of distinct transaction partners over the lifetime of the network. Both of these distributions are fat-tailed and are robust over the period of almost ten years despite the size of the network increasing by multiple orders of magnitude. The black line in the left figure shows a power-law fit for the final distribution that has an exponent of $2.68$. The fit was carried out with the \texttt{plfit} package~\cite{plfit}, based on the algorithm of Clauset~et~al.~\cite{clauset}.}
	\label{bitcoin_dist}
\end{figure*}

\begin{figure*}
	\includegraphics{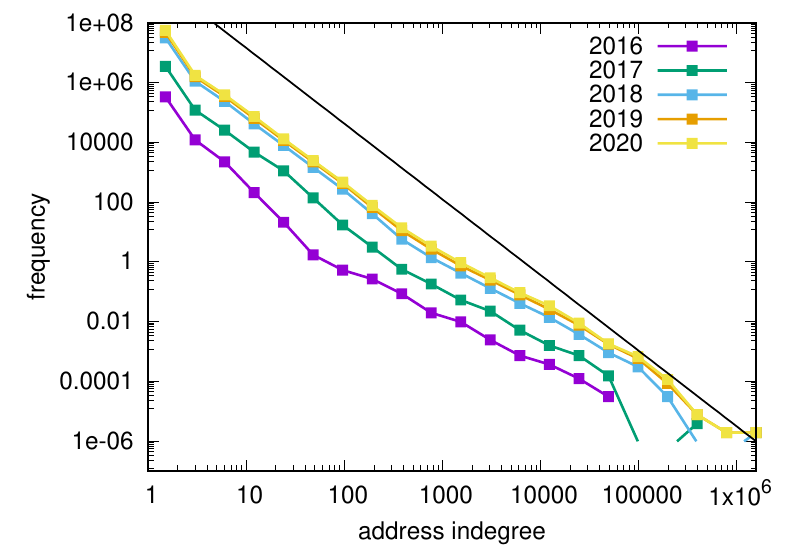}
	\includegraphics{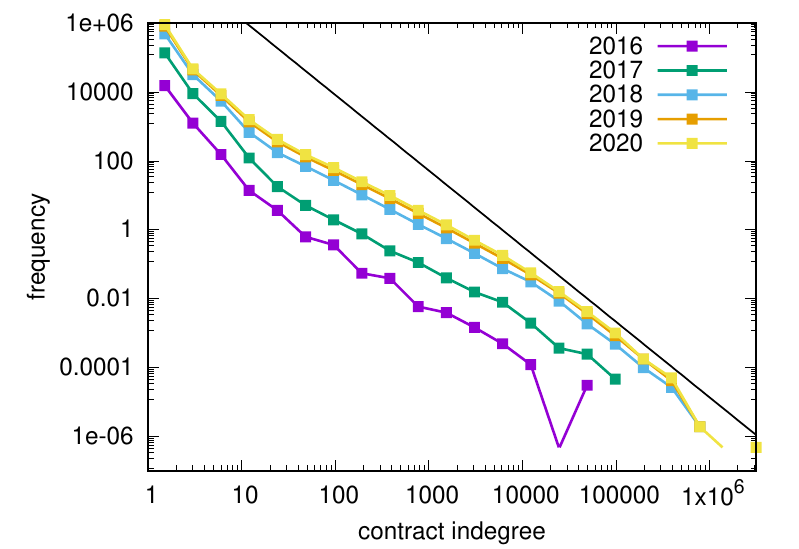}
	\caption{Indegree distribution of regular addresses (left) and contract addresses (right) in Ethereum. These distributions are also characterized as fat-tailed ones, and are well approximated by power-laws, similarly to Bitcoin. Again, the time evolution is robust over a period of almost five years, during which the Ethereum network grew over 100-fold. Black lines show power-law fits for the final distribution, with exponents of $-2.54$ and $-2.19$ for addresses and contracts respectively. Fits were carried out with the \texttt{plfit} package~\cite{plfit}, based on the algorithm of Clauset~et~al.~\cite{clauset}.}
	\label{eth_dist}
\end{figure*}

We perform a simple characterization of structure by looking at the degree distribution of transaction networks. More specifically, we are interested in indegree distributions, since this can be interpreted as a measure of capacity to attract interaction with external entities. Both networks are characterized by fat-tailed distributions over their lifetime that are well approximated with power-laws (Figs.~\ref{bitcoin_dist} and~\ref{eth_dist}). The stability in shape of these distributions is especially remarkable considering that different stages of the networks depicted in Figs.~\ref{bitcoin_dist} and~\ref{eth_dist} represent an over 100-fold increase in size (over 10,000-fold increase in the case of Bitcoin when comparing very early instances with the latest ones).

\subsection{Preferential attachment}

\begin{figure*}
	\centering
	\includegraphics{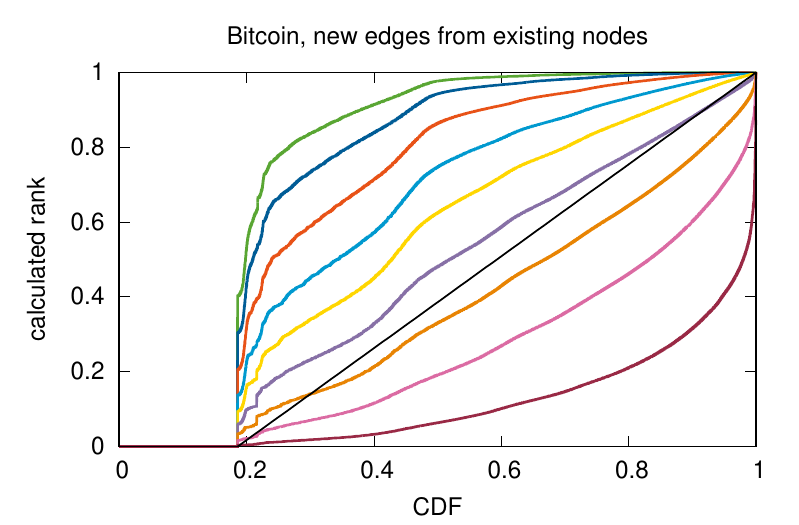}
	\includegraphics{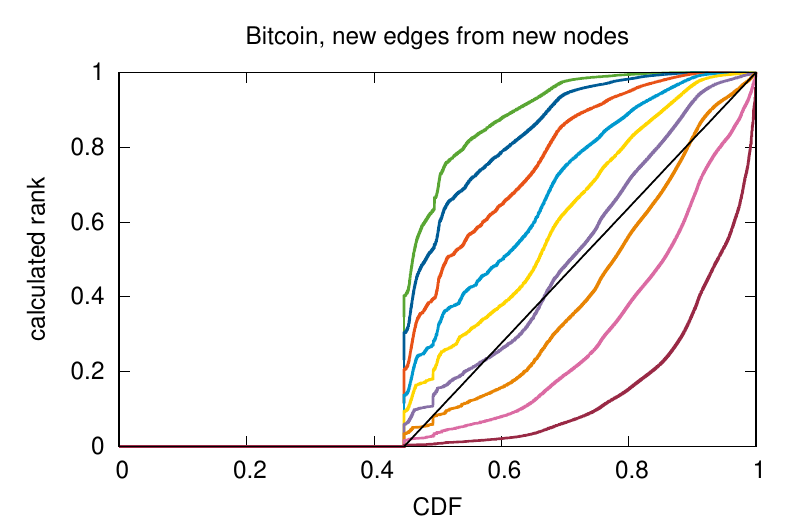} \\
	\includegraphics{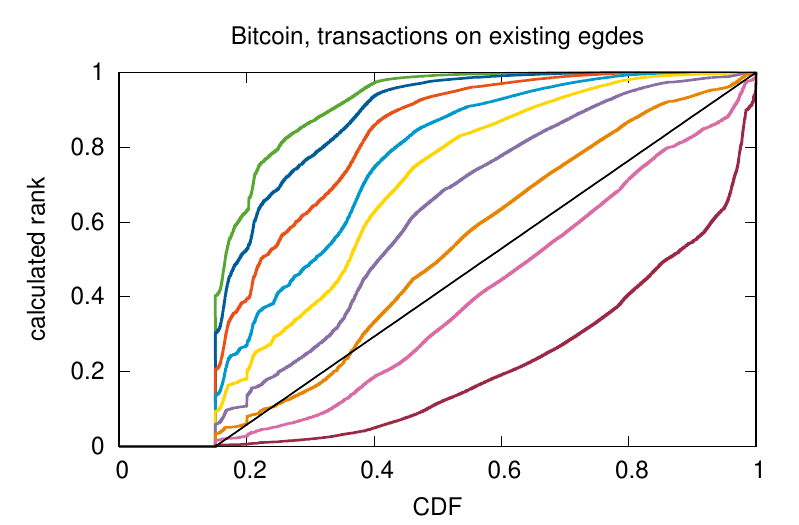}
	\includegraphics{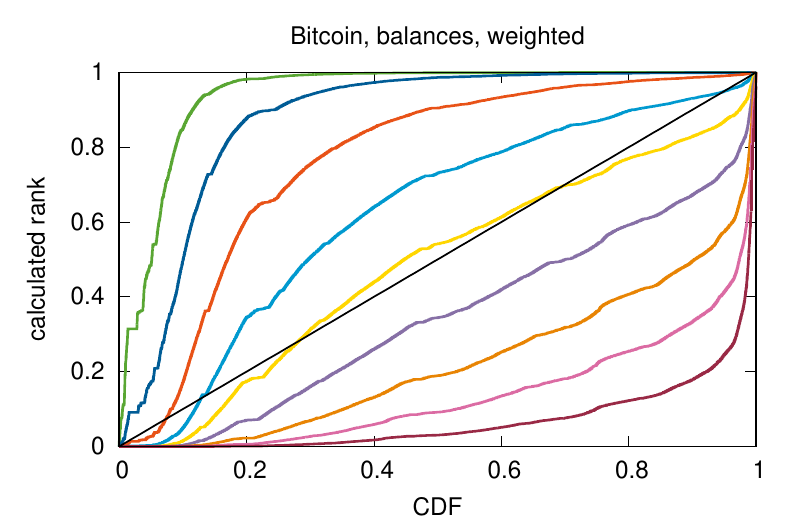} \\
	\includegraphics{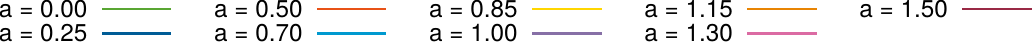}
	\caption{Testing for preferential attachment in Bitcoin. The four panels show the cumulative distribution of transformed ranks in the case of four different types of events. All cases exhibit a clear sign of preferential attachment. At the same time, there is a clear case of transactions that target new nodes (i.e.~nodes with zero degree). This is understandable given the nature of Bitcoin, where users are encouraged to frequently generate new addresses to enhance privacy. Black lines show the expected ideal (i.e.~uniform) distribution. Kolmogorov-Smirnov differences from these distributions are shown in Fig.~\ref{res:bitcoin_exp}.}
	\label{res:bitcoin}
\end{figure*}

\begin{figure*}
	\centering
	\includegraphics{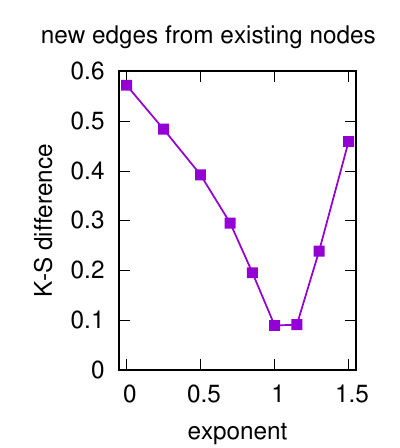}
	\includegraphics{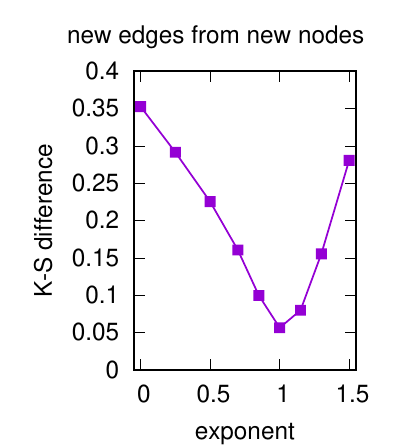}
	\includegraphics{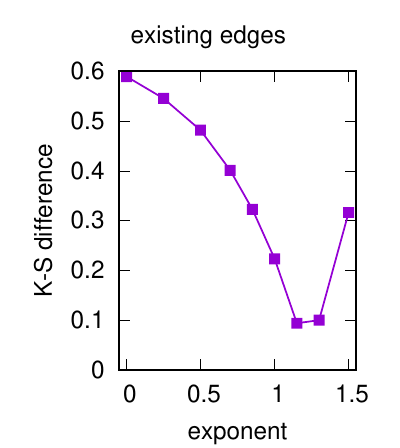}
	\includegraphics{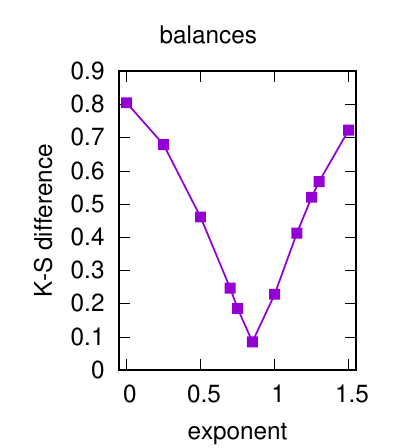}
	\caption{Kolmogorov-Smirnov differences from the presumed uniform distribution for the case of preferential attachment in Bitcoin, i.e.~for results displayed in Fig.~\ref{res:bitcoin}.}
	\label{res:bitcoin_exp}
\end{figure*}

\begin{figure*}
	\centering
	\includegraphics{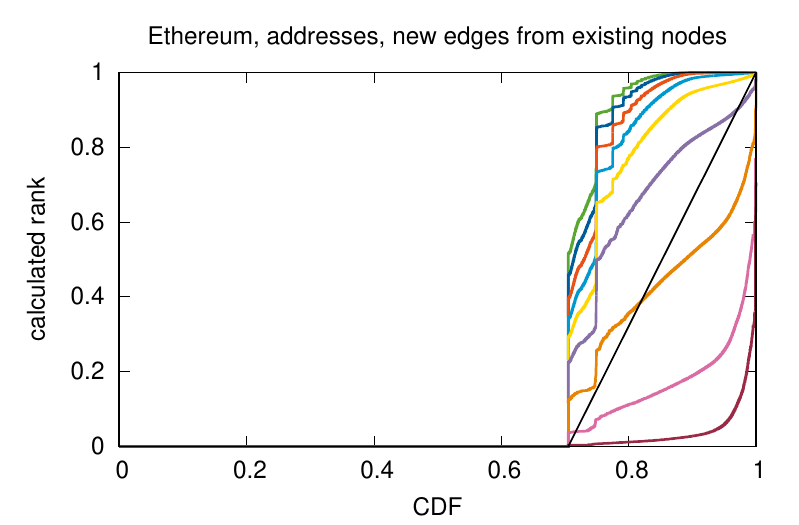}
	\includegraphics{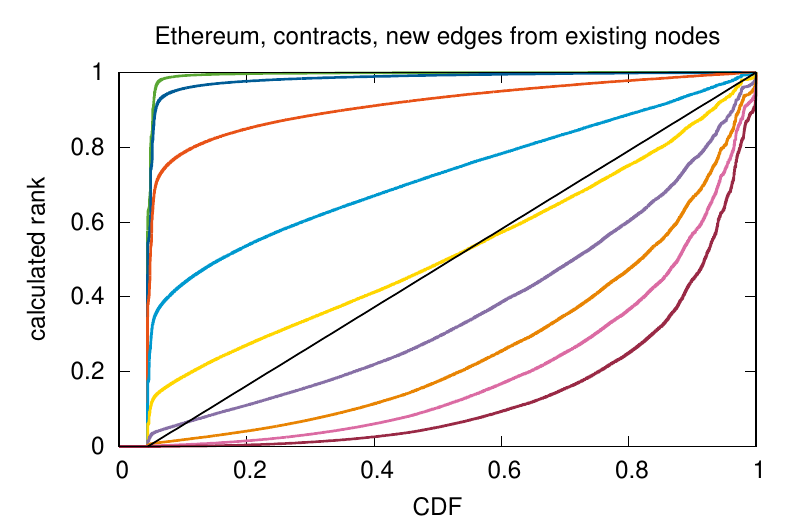} \\
	\includegraphics{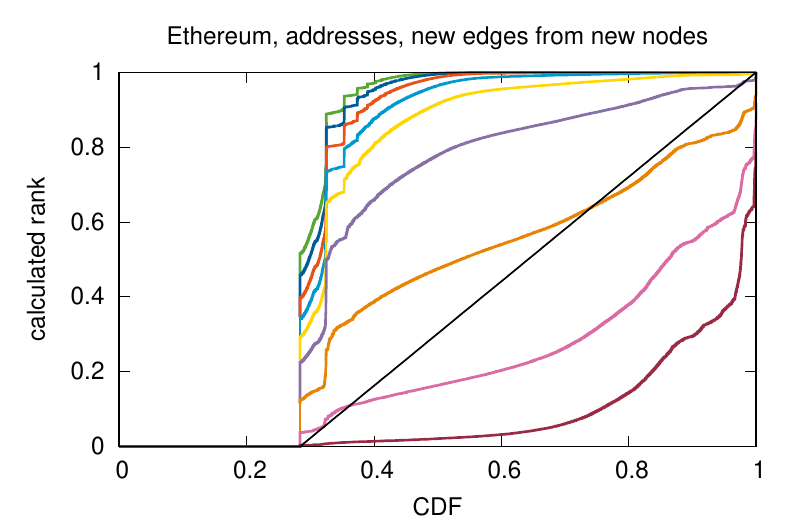}
	\includegraphics{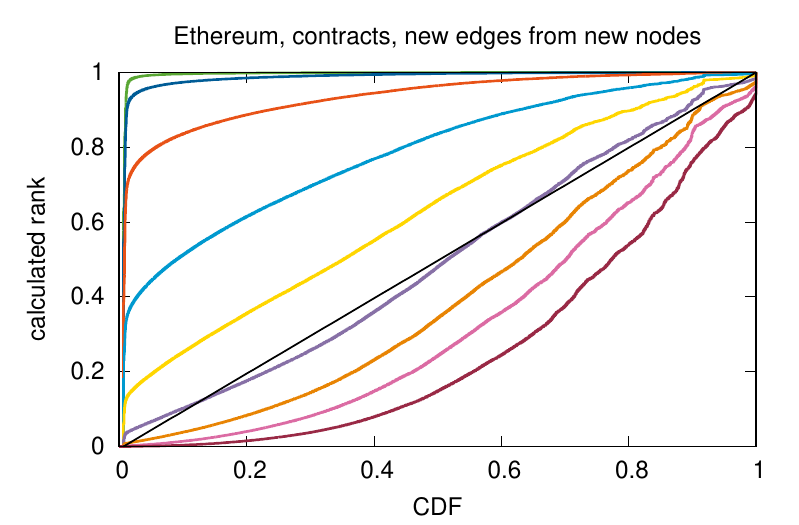} \\
	\includegraphics{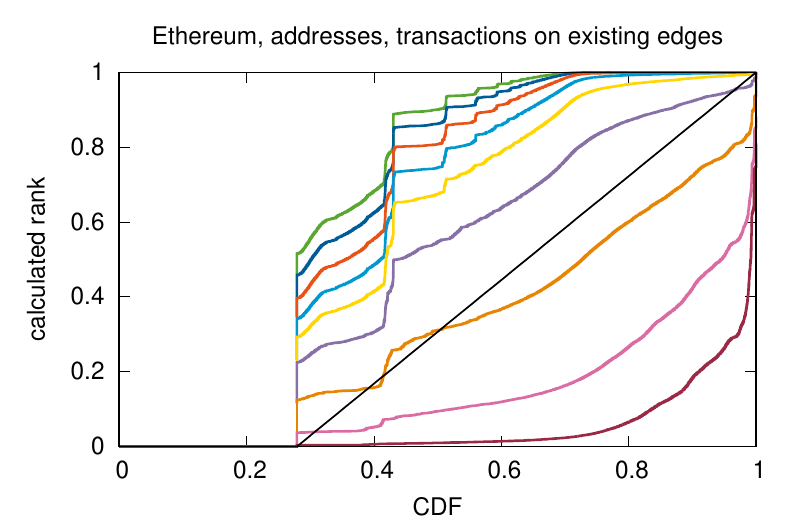}
	\includegraphics{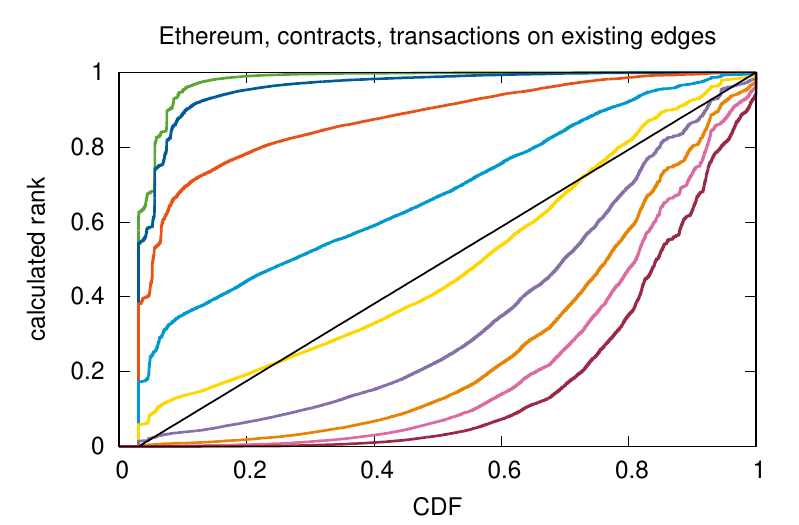} \\
	\includegraphics{figs_bal_legend}
	\caption{Testing for preferential attachment in Ethereum. The left column shows edges where the target is a regular address, while the right column shows edges where the target is a smart contract. Black lines show the expected ideal (i.e.~uniform) distribution. Kolmogorov-Smirnov differences from these distributions are shown in Fig.~\ref{res:ethereum_exp}.}
	\label{res:ethereum}
\end{figure*}

We test for the presence of preferential attachment by considering all transactions that add new links to the aggregated networks and calculating transformed ranks according to Eq.~(\ref{eq:R}). In Fig.~\ref{res:bitcoin} and Fig.~\ref{res:ethereum}, we display the transformed ranks in order, i.e.~as a function of their cumulative distribution function (CDF), for the case of the Bitcoin and Ethereum transaction networks, and for the evolution of Bitcoin balances. For each case, a perfect fit with the model of nonlinear preferential attachment (i.e.~Eq.~(\ref{eq:model})) would be a straight line, corresponding to the case where the transformed ranks are uniformly distributed in the $[0,1]$ interval. Finding an exponent that best describes the process means finding a case where a straight line best approximates the distribution of transformed rank values.

In most cases, a significant feature is that the distributions do not start from zero. The means that there is a large number of transactions that target newly created addresses, in contrast to the original nonlinear preferential attachment model, where the probability of an edge targeting a non-existent node (i.e.~a node with a degree of zero) is zero. This is understandable given that users can freely create any number of addresses, and are advised to often move their wealth to new addresses. Also, many service providers create unique addresses for their customers, which necessarily have zero degree then. Given this, we might need to restrict the preferential attachment model to only apply to existing addresses, while we acknowledge that linking to new addresses is governed by more specific rules that are relevant to cryptocurrency system usage.

\begin{figure*}
	\centering
	\includegraphics{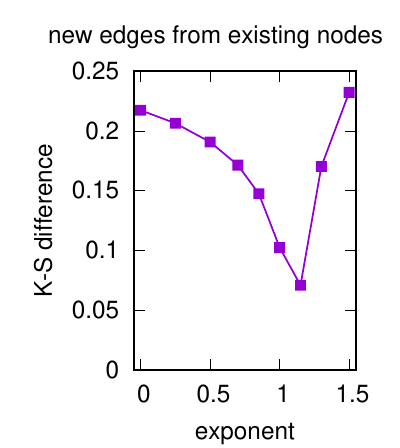}
	\includegraphics{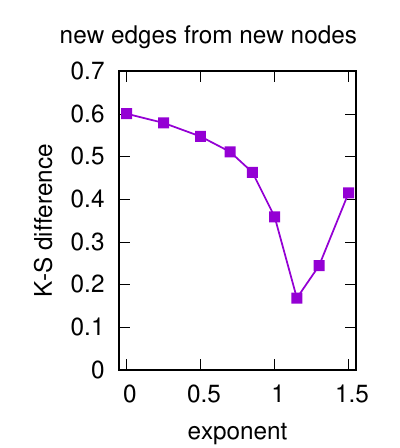}
	\includegraphics{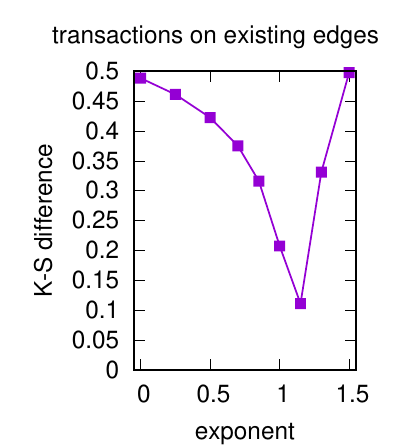}\\
	\includegraphics{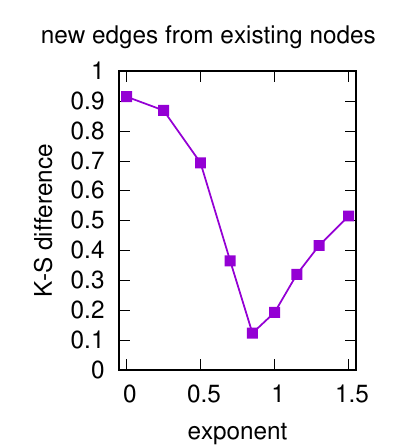}
	\includegraphics{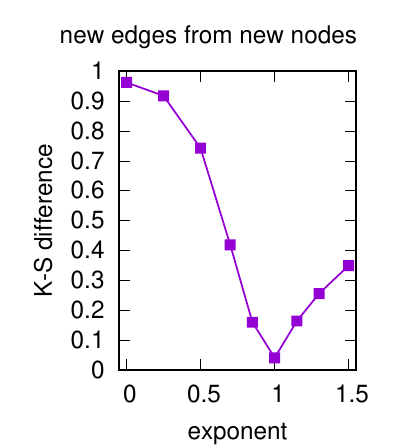}
	\includegraphics{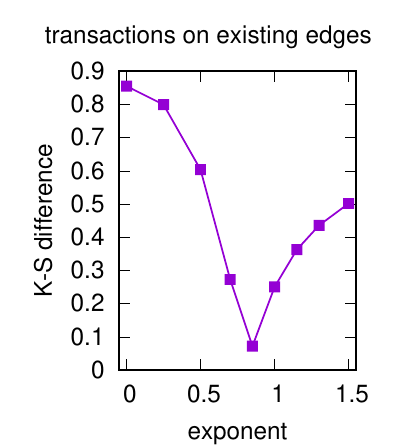}
	\caption{Kolmogorov-Smirnov differences from the presumed uniform distribution for the case of preferential attachment in Ethereum, i.e.~for results displayed in Fig.~\ref{res:ethereum}. Top row: results for transactions targeting addresses; bottom row: results for transactions targeting contracts.}
	\label{res:ethereum_exp}
\end{figure*}

Given this observation, we only focus on nonzero transformed ranks when considering if they can be fitted with a uniform distribution. Graphically, this corresponds to starting the lines that represent such uniform distributions (the black lines in Figs.~\ref{res:bitcoin} and~\ref{res:ethereum}) from the CDF value that corresponds to the first nonzero transformed rank.

In each case, we see strong evidence for the presence of a preferential attachment process. This is clear by the fact that ranks calculated under the $\alpha = 0$ assumption always result in a much worse fit than $\alpha > 0$ exponents. Beside visual inspection of the fits, we calculate the Kolmogorov-Smirnov difference from the presumed uniform distributions, and present this as a function of the $\alpha$ presumed exponent in Figs.~\ref{res:bitcoin_exp} and~\ref{res:ethereum_exp}. Overall, exponents around $\alpha = 1$ give the best fits; but there are some further interesting observations regarding typical values.

In the case of the Bitcoin transaction network, linear preferential attachment is the most plausible model for the case of newly created edges, either from new or from existing nodes. This is consistent with our earlier results~\cite{Kondor2014b} that were done for this network at a much earlier stage. For the case of repeated edges (i.e.~repeated transactions on edges that appeared before), we see a slight superlinear case, with $\alpha = 1.15$ and $\alpha = 1.3$ both giving almost equally plausible fits. Furthermore, we also tested for preferential attachment in the case of money dynamics, i.e.~related to the flow of Bitcoins. In this case, instead of node degrees, we considered the balance of the target address, and also weighted the CDF values with the transferred Bitcoin amount. In this case, we see evidence of slightly sublinear preferential attachment, with $\alpha = 0.85$ being the most plausible exponent. This is again consistent with our earlier results~\cite{Kondor2014b}.

In the case of Ethereum, we separately analyze the case where edges connect to regular addresses (left column in Fig.~\ref{res:ethereum}; top row in Fig.~\ref{res:ethereum_exp}) and the case where the target of an edge is a smart contract (right column in Fig.~\ref{res:ethereum}; bottom row in Fig.~\ref{res:ethereum_exp}). For regular addresses, we see some evidence of superlinear preferential attachment ($\alpha = 1.15$ being the most plausible exponent); nevertheless, a uniform distribution does not seem a very good fit in this case, as we see significant further features in the distribution of transformed ranks in Fig.~\ref{res:ethereum}. Still, we can say that a form of preferential attachment is important in this process, since the case of $\alpha = 0$ gives a much worse agreement with the empirical distribution of transformed ranks than any other case. For smart contracts, the distributions fit more nicely, and suggest a slightly sublinear process, with $\alpha = 0.85$ being the most plausible exponent, with the exception of the case, where a newly created address initiates a transaction; in this case, $\alpha = 1$ gives better fit.

\subsubsection{Limited lifetime edges}

We repeated the procedure of calculating the transformed ranks for variants of the transaction networks where edges are assumed to have limited lifetimes, i.e.~one day or 30 days. This means that indegrees of nodes can decrease in the case when edges are removed. Detailed results are shown in the Supplementary Material, in Figs.~\ref{bitcoin2}--\ref{ethereum_30d_exp}. These results are highly consistent with what we have obtained for the fully time aggregated network, showing an evidence of preferential attachment as well. Best fitting exponents are very similar in all cases for Bitcoin, while for Ethereum addresses, we see slightly higher exponents for short time intervals, hinting at a preference for addresses that already were the target of high activity recently.

\subsubsection{Evaluating changes in exponents over time}

\begin{figure*}
\centering
\includegraphics{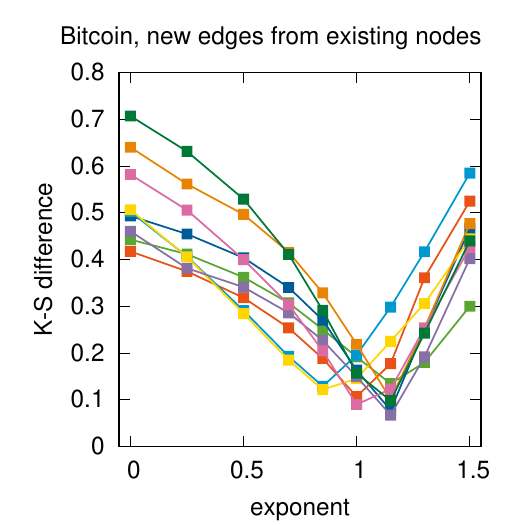}
\includegraphics{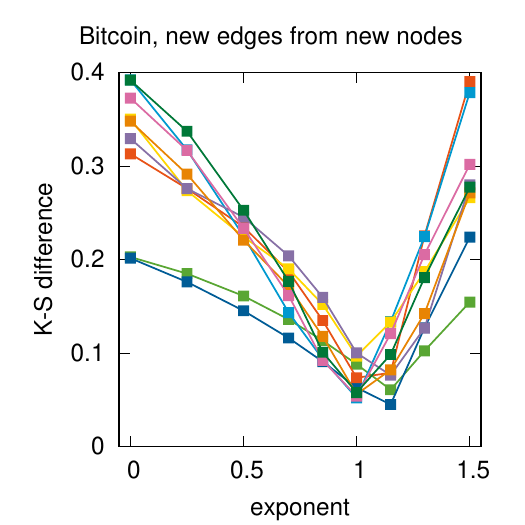}
\includegraphics{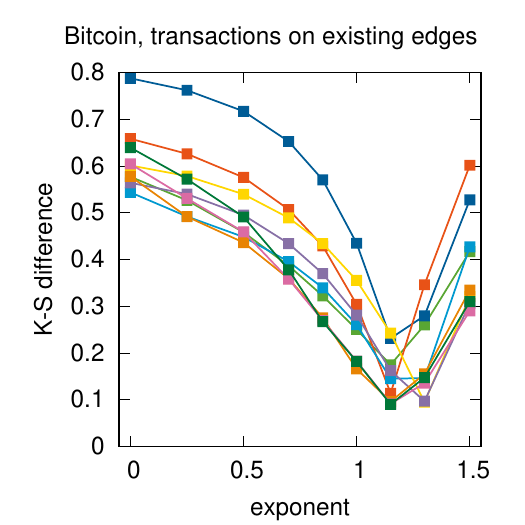}
\includegraphics{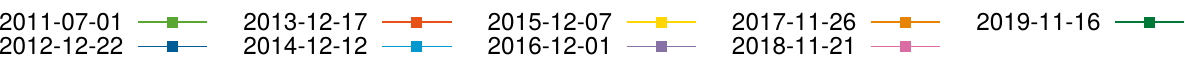}
\caption{Kolmogorov-Smirnov differences from the presumed uniform distribution for the case of preferential attachment in Bitcoin, for distributions disaggregated over time. Each line corresponds to a distribution that was compiled based on the events taking place in the six month prior to it.}
\label{res:bitcoin_exp_multi}
\end{figure*}

\begin{figure*}
\centering
\includegraphics{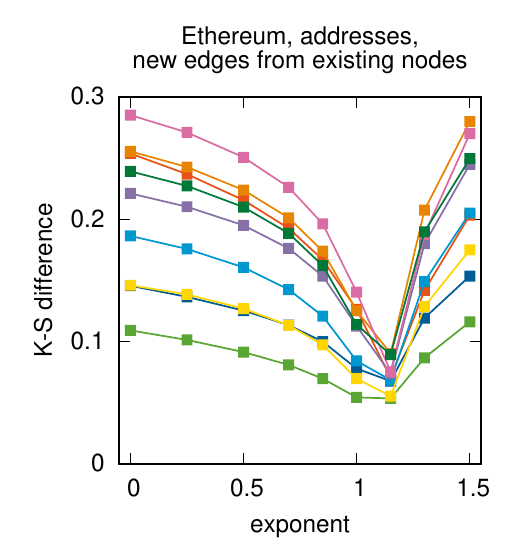}
\includegraphics{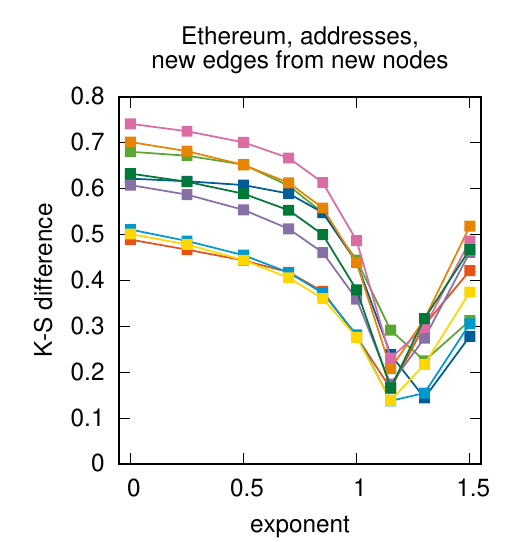}
\includegraphics{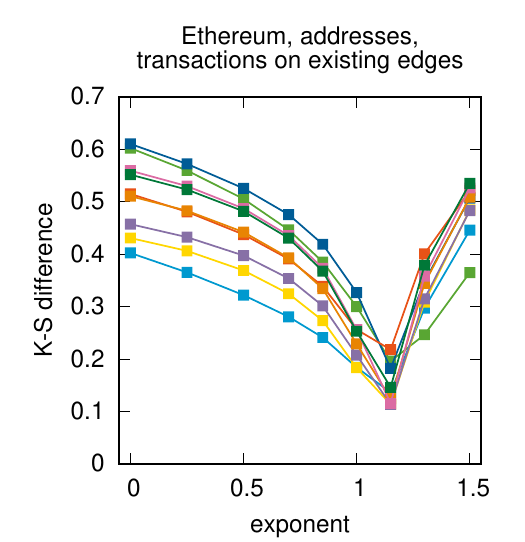}
\includegraphics{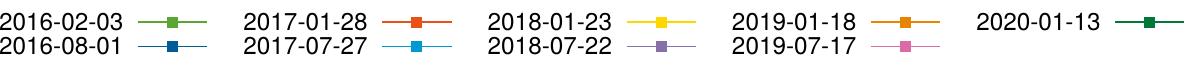}
\caption{Kolmogorov-Smirnov differences from the presumed uniform distribution for the case of preferential attachment in Ethereum, for transactions targeting regular addresses, distributions disaggregated over time. Each line corresponds to a distribution that was compiled based on the events taking place in the six month prior to it.}
\label{res:ethereum_addresses_exp_multi}
\end{figure*}

\begin{figure*}
\centering
\includegraphics{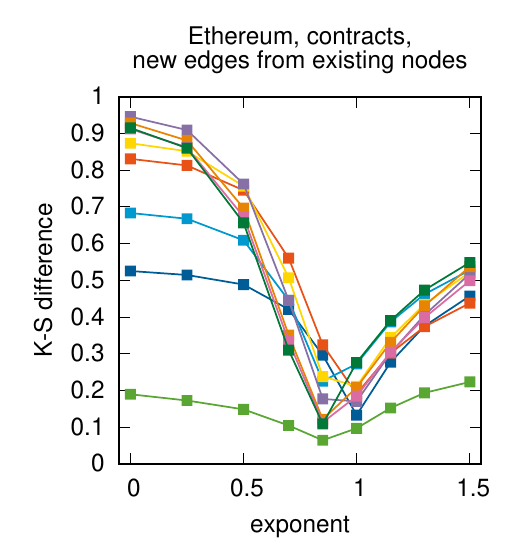}
\includegraphics{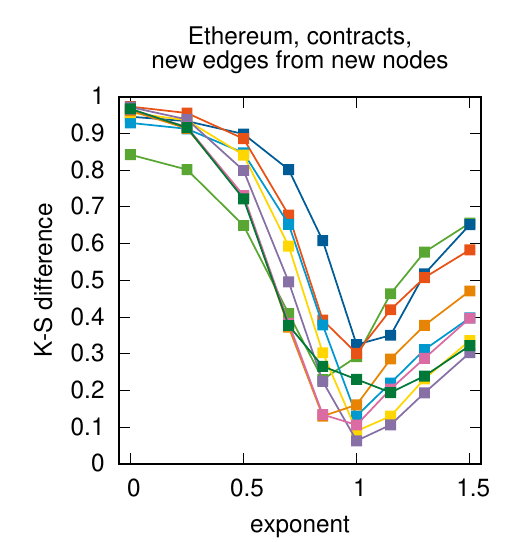}
\includegraphics{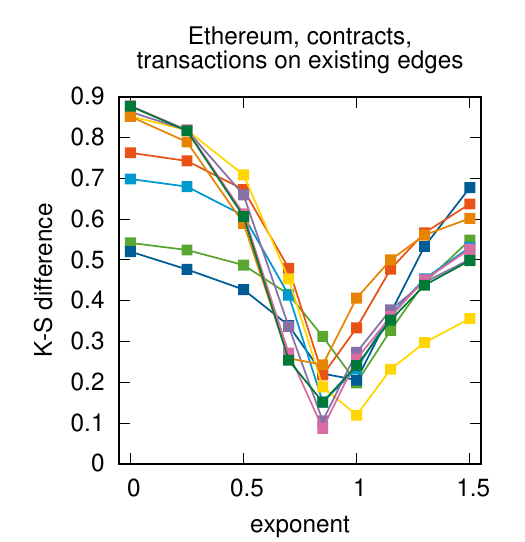}
\includegraphics{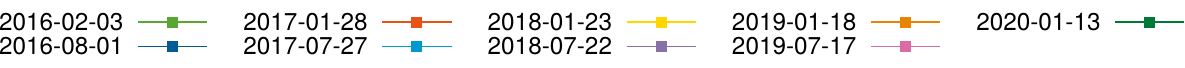}
\caption{Kolmogorov-Smirnov differences from the presumed uniform distribution for the case of preferential attachment in Ethereum, for transactions targeting smart contracts, distributions disaggregated over time. Each line corresponds to a distribution that was compiled based on the events taking place in the six month prior to it.}
\label{res:ethereum_contracts_exp_multi}
\end{figure*}

So far, we have evaluated statistics of preferential attachment in a time-aggregated fashion, i.e.~we considered all transaction that happened over their lifetime when looking at the distribution of transformed ranks. To gain more insights into the process of network evolution, we evaluated the distribution of transformed ranks in shorter, half-year long time intervals, and show the Kolmogorov-Smirnov distances as a function of exponents in Figs.~\ref{res:bitcoin_exp_multi}, \ref{res:ethereum_addresses_exp_multi} and~\ref{res:ethereum_contracts_exp_multi}. We see that while the best fit is achieved around the typical value of exponents as found previously (see Figs.~\ref{res:bitcoin_exp} and~\ref{res:ethereum_exp}), there is some noticeable variation, with some time periods showing slightly smaller or larger exponents as best fits. This hints that there might be important time-dependent processes shaping shaping the evolution of the transaction networks beyond preferential attachment, as also evidences by the deviations of the perfect fit of the transformed rank distributions.

\section{Discussion}

Our results confirm that preferential attachment is a key component shaping the evolution of cryptocurrency transaction networks, contributing to the heavy-tailed degree distributions that arise. This is true regardless of the time scale considered, as focusing only on the subnetworks of recent transaction partners results in very similar statistics of edge creation and activity. While our previous results showed the presence of preferential attachment in the early Bitcoin network, it is remarkable that the same dynamic is present over a much longer time period that involved an almost 100-fold growth in terms of network size and several up- and downturns in the market.

Findings of preferential attachment and heavy-tailed degree distributions matches well with other findings about networks that describe interactions between complex and self-organizing social, technological or economical phenomena. It is also consistent with the picture of cryptocurrency networks being made up of a few very large players interacting with regular users who have limited activity, especially when considered on the level of individual addresses.

Our work suggests several future directions for research. Firstly, while we find that preferential attachment is consistently present in all of the studied networks over their lifetime, our results hint that the detailed dynamics of the process (as represented by the best fitting exponent, and also the shape of the distribution of transformed ranks) changes over time (see Figs.~\ref{res:bitcoin_exp_multi}, \ref{res:ethereum_addresses_exp_multi} and~\ref{res:ethereum_contracts_exp_multi}). A more in-depth investigation of these changes could lead to new insights about different phases of cryptocurrency usage and how it is linked to structural properties of the transaction network.

Second, while the overall trend of preferential attachment is quite clear, there are systematic deviations from a perfect fit to the presumed form (Eq.~(\ref{eq:model})). It is a question whether these could be explained by modifying the functional form or extending it to include readily available properties of nodes. Research in this direction could uncover more detailed driving forces of transaction network evolution and provide new, generalizable models of network growth~\cite{Naglic2019}.

Finally, depending on availability of datasets, a comparison between cryptocurrencies and other types of economical or financial transaction networks could inform about the generalizability of our findings and also help in better understanding the role cryptocurrencies play in the global economy~\cite{Alabi2017,Seebacher2018}, a still widely debated subject. To facilitate further research, we publish the data and code used in the current work~\cite{dataset, ethdata, orbtree, patest_new}.

\section*{Conflict of Interest Statement}

The authors declare that the research was conducted in the absence of any commercial or financial relationships that could be construed as a potential conflict of interest.

\section*{Author Contributions}

DK, GV, IC contributed to conception and design of the study. DK contributed software. NB, JS performed data collection and preprocessing. DK, NB, JS analyzed data. DK performed further data analysis and drafted the paper. All authors contributed to manuscript revision, read, and approved the submitted version.

\section*{Acknowledgements}

This research was supported by the Hungarian Ministry of Innovation and Technology and the National Research, Development and Innovation Office within the Quantum Information National Laboratory of Hungary.

This research is supported by the Singapore Ministry of National Development and the National Research Foundation, Prime Minister’s Office, under the Singapore-MIT Alliance for Research and Technology (SMART) programme.

\section*{Data Availability Statement}
The datasets for this study are made available as Refs.~\cite{dataset} and~\cite{ethdata}. Code and scripts used to generate the main results of this paper are available as Ref.~\cite{patest_new}.

\bibliographystyle{unsrtnat}

\newpage
\normalsize

\twocolumn[  
    \begin{@twocolumnfalse}
    
		\begin{center}
			{\huge \bf Supplementary Material \\[2.5ex]}
		\end{center}
		
    \end{@twocolumnfalse}
]

\renewcommand{\thefigure}{S\arabic{figure}}
\renewcommand{\thetable}{S\arabic{table}}
\renewcommand{\thealgorithm}{S\arabic{algorithm}}

\setcounter{figure}{0}
\setcounter{table}{0}
\setcounter{algorithm}{0}

\makeatletter
\let\ftype@table\ftype@figure
\makeatother

\section{Edges with limited lifetime}

Beside the analysis presented in the main text, we evaluated preferential attachment statistics in two additional cases, where we assume that each edge in the networks has a limited ``lifetime'': it is erased after a given period of time if activity is not repeated on it. In practice, we repeated our main analysis with presumed lifetimes of one and 30 days. The former corresponds to a case where we assume that linking preference is related to the incoming transactions an address had on the previous day, while the latter assumes that transactions in the past month are considered.

\begin{figure*}
	\centering
	\large One day edge lifetime \quad \quad \quad \quad \quad \quad \quad \quad \quad 30 day edge lifetime \\
	\includegraphics{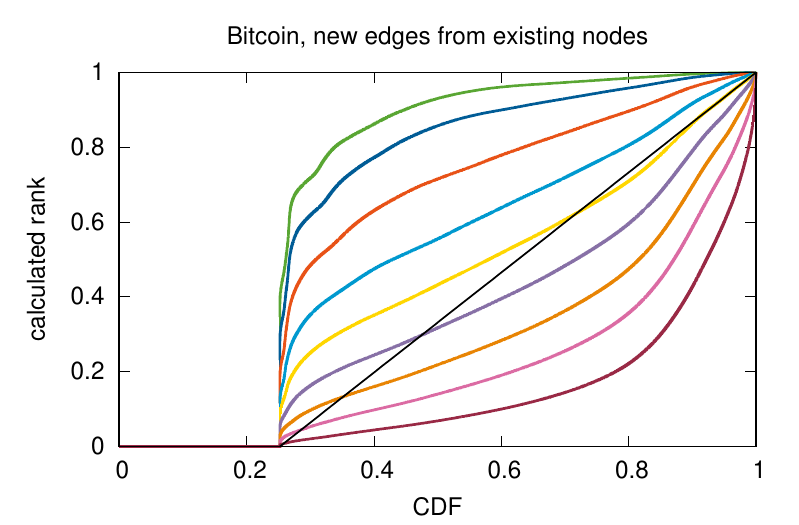} \includegraphics{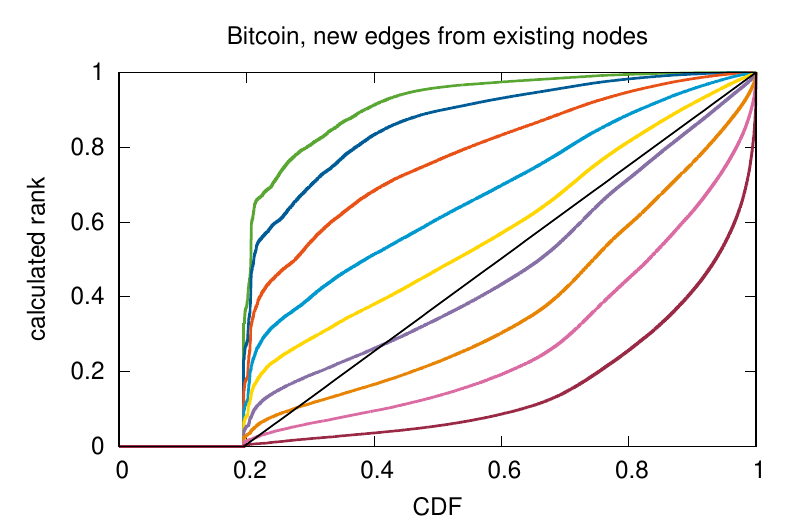} \\
	\includegraphics{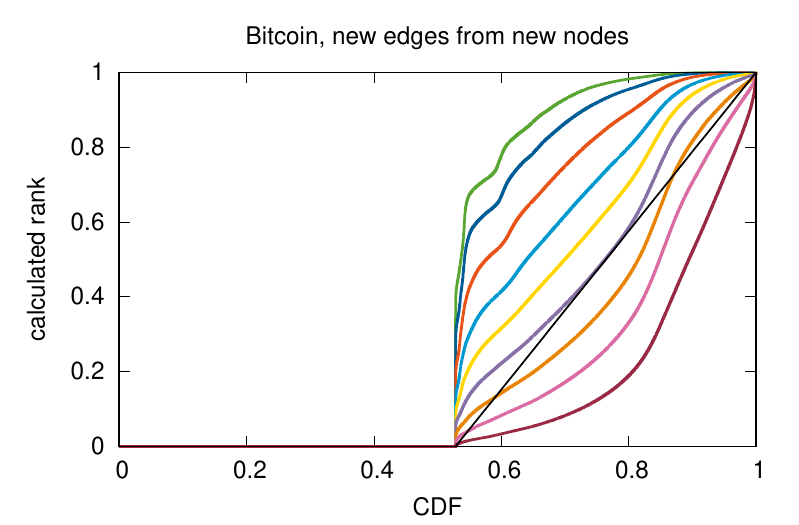} \includegraphics{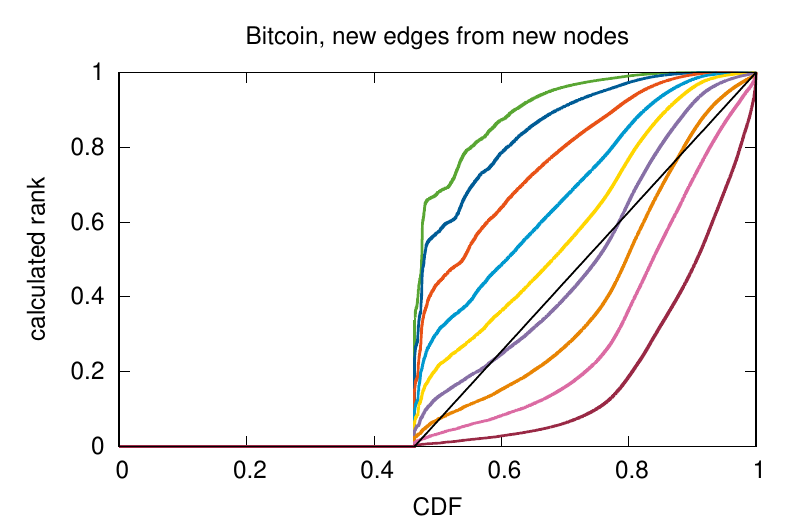} \\
	\includegraphics{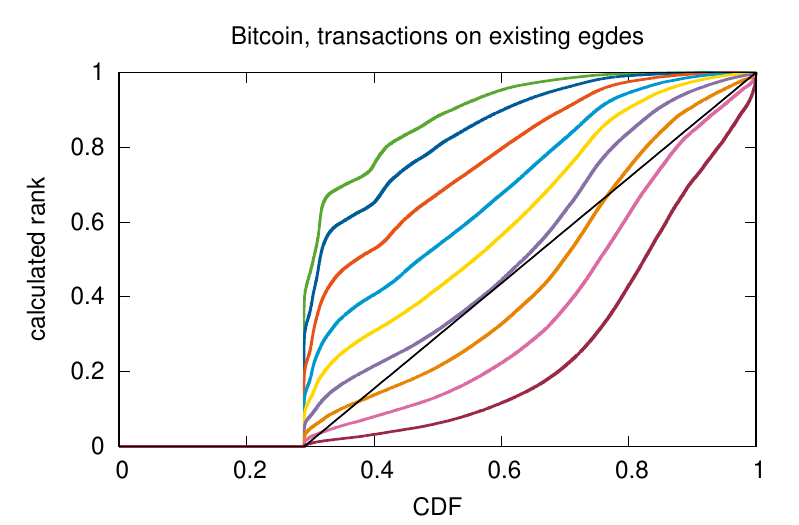} \includegraphics{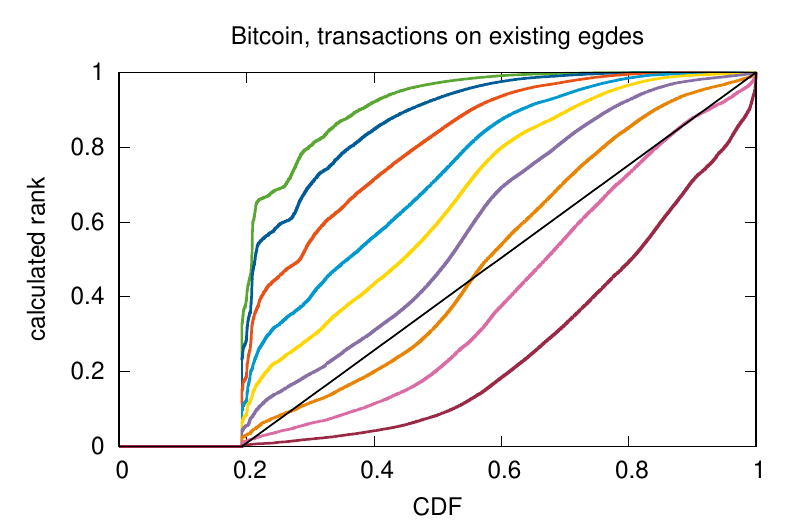} \\
	\includegraphics{figs_bal_legend}
	\caption{Testing for preferential attachment in Bitcoin, assuming a one day (left column) or a 30 day ``lifetime'' for edges (right column). Three rows show the cumulative distribution of transformed ranks in the case of three different types of events, all of which exhibit preferential attachment in a very similar fashion to the full-time aggregated networks. Black lines show the expected ideal (i.e.~uniform) distribution. Kolmogorov-Smirnov differences from these distributions are shown in Figs.~\ref{bitcoin_exp_1d} and~\ref{bitcoin_exp_30d}.}
	\label{bitcoin2}
\end{figure*}

\begin{figure*}
	\centering
	\includegraphics{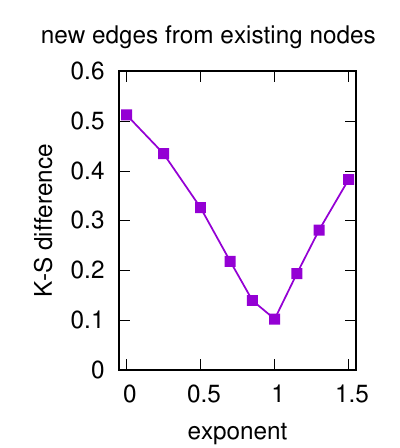}
	\includegraphics{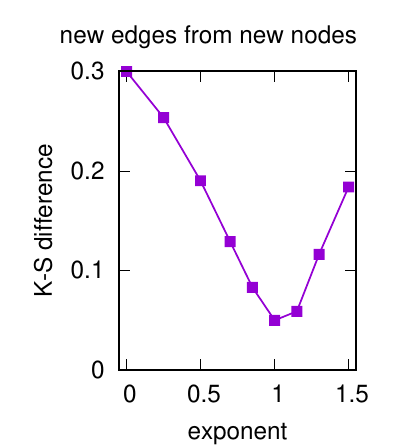}
	\includegraphics{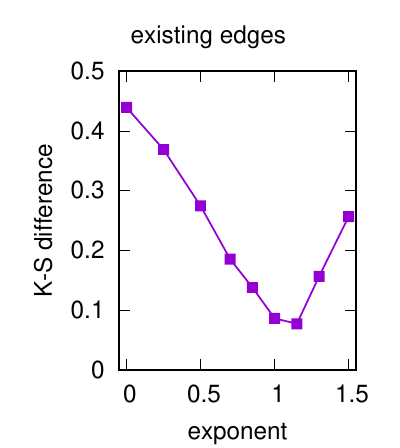}
	\caption{Kolmogorov-Smirnov differences from the presumed uniform distribution for the case of preferential attachment in Bitcoin, assuming one day ``lifetime'' of edges, i.e.~for results displayed in Fig.~\ref{bitcoin2}.}
	\label{bitcoin_exp_1d}
\end{figure*}

\begin{figure*}
	\centering
	\includegraphics{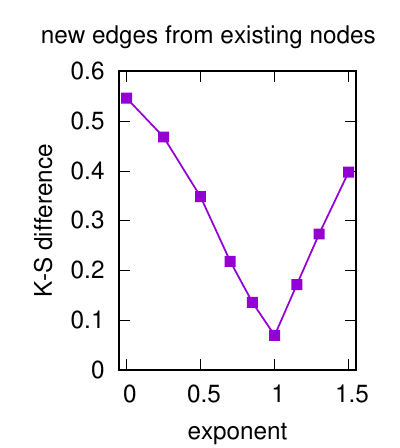}
	\includegraphics{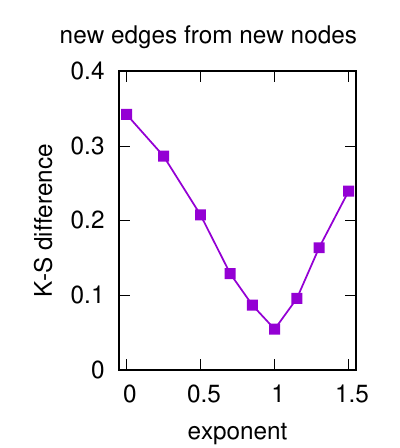}
	\includegraphics{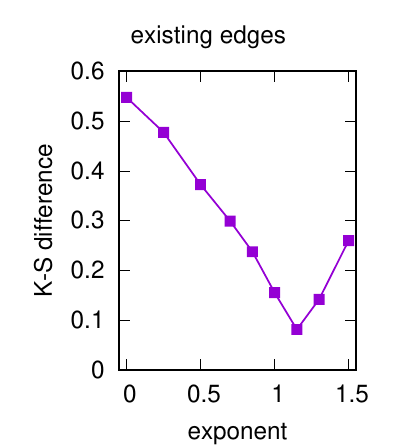}
	\caption{Kolmogorov-Smirnov differences from the presumed uniform distribution for the case of preferential attachment in Bitcoin, assuming a 30 day ``lifetime'' of edges, i.e.~for results displayed in Fig.~\ref{bitcoin2}.}
	\label{bitcoin_exp_30d}
\end{figure*}

\begin{figure*}
	\centering
	\large Ethereum, one day edge lifetime \\
	\includegraphics{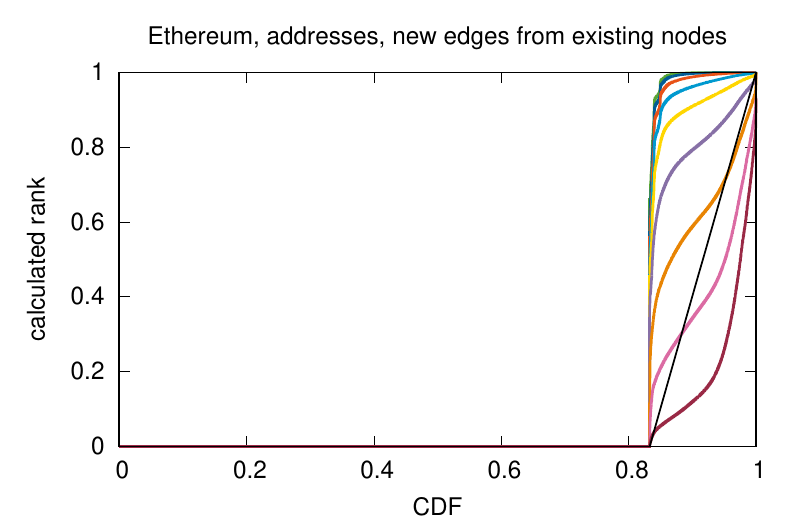}
	\includegraphics{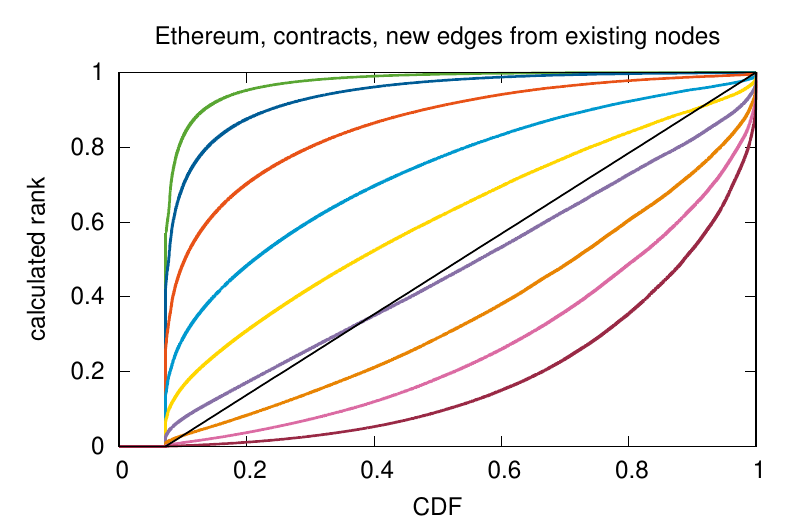} \\
	\includegraphics{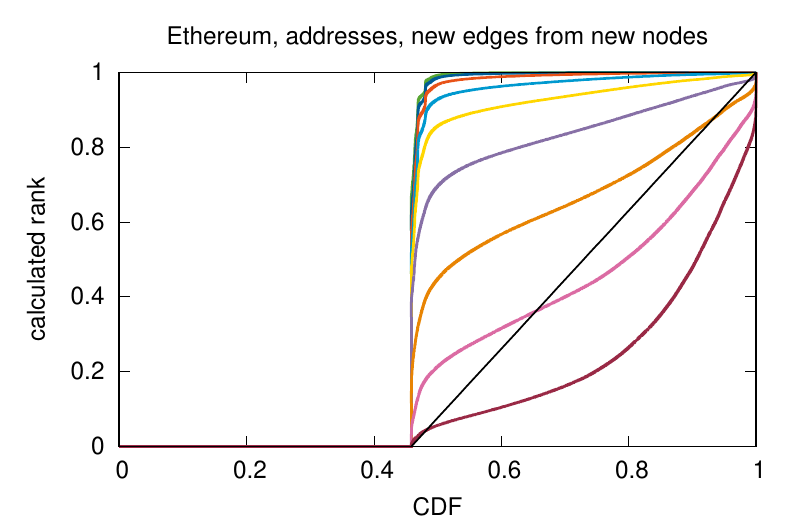}
	\includegraphics{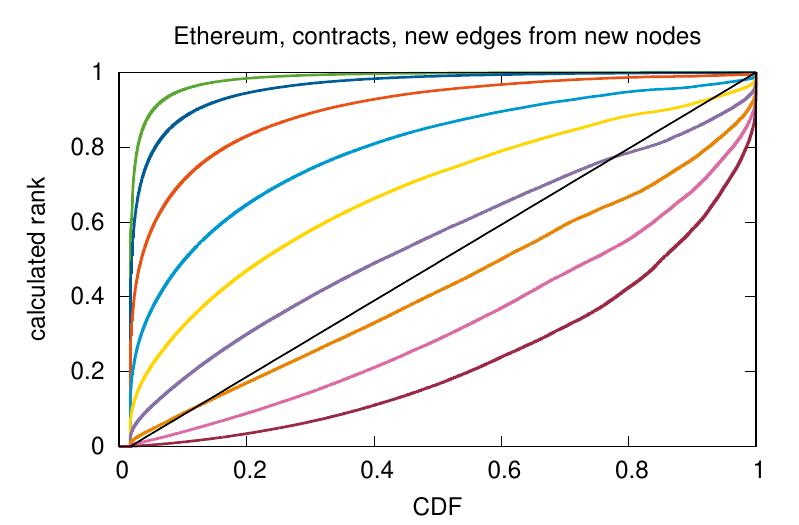} \\
	\includegraphics{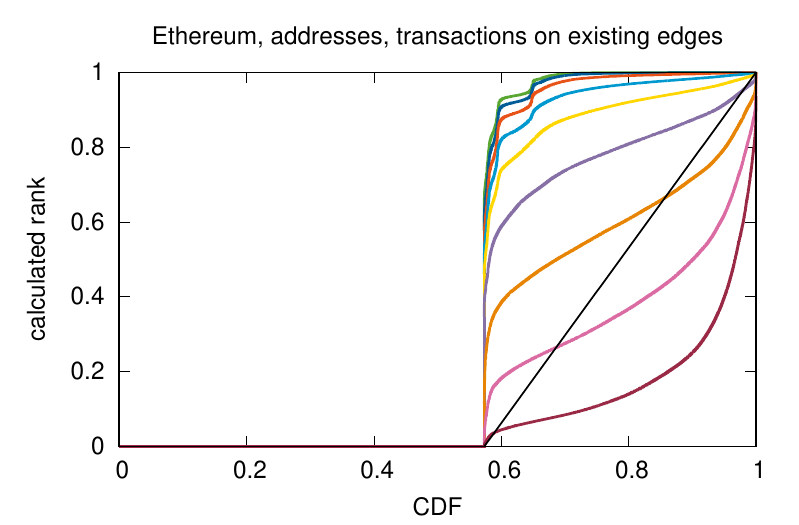}
	\includegraphics{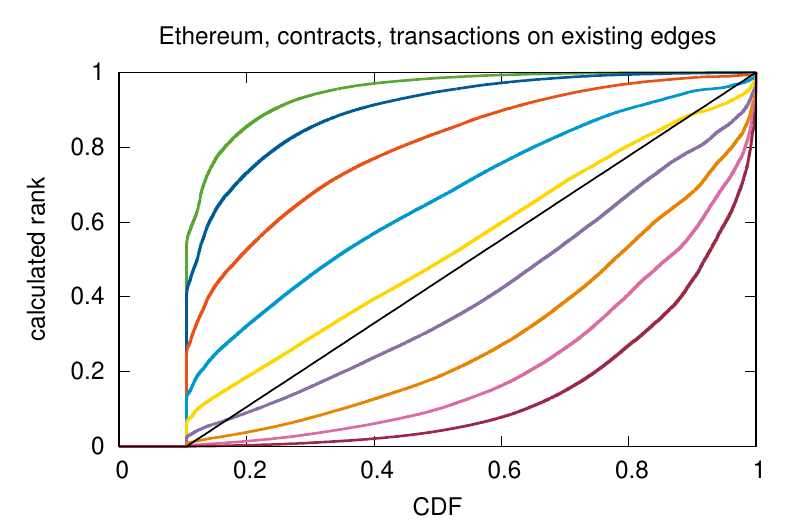} \\
	\includegraphics{figs_bal_legend}
	\caption{Testing for preferential attachment in Ethereum, assuming a one day ``lifetime'' for edges. The left column shows edges where the target is a regular address, while the right column shows edges where the target is a smart contract. Black lines show the expected ideal (i.e.~uniform) distribution. Kolmogorov-Smirnov differences from these distributions are shown in Fig.~\ref{ethereum_1d_exp}.}
	\label{ethereum_1d}
\end{figure*}

\begin{figure*}
	\centering
	\large Ethereum, 30 day edge lifetime \\
	\includegraphics{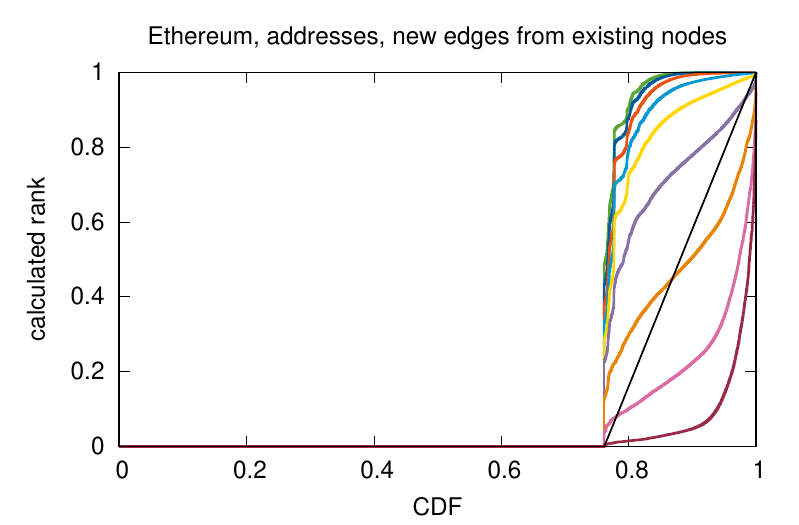}
	\includegraphics{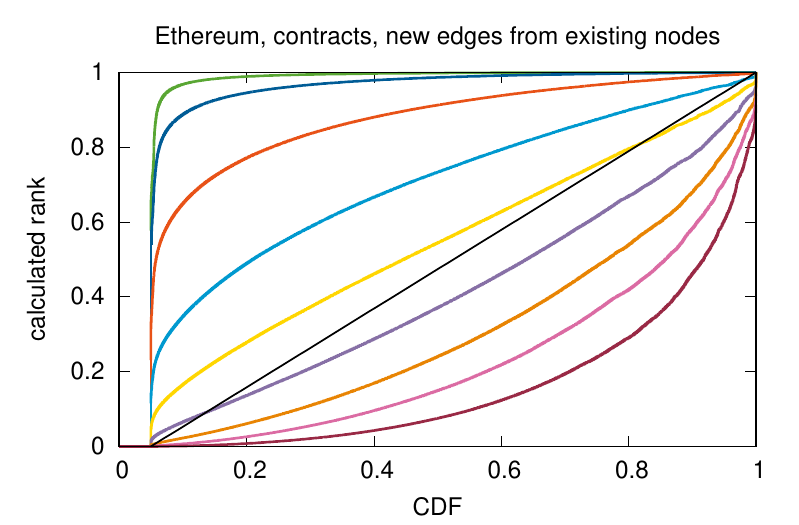} \\
	\includegraphics{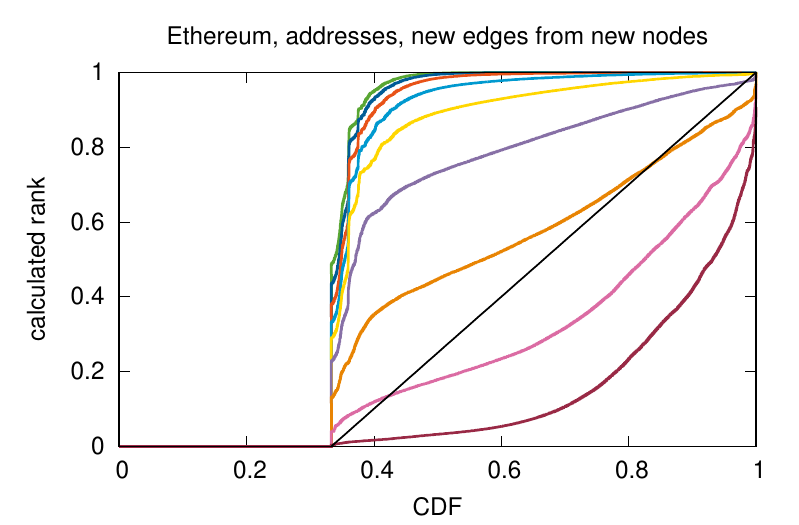}
	\includegraphics{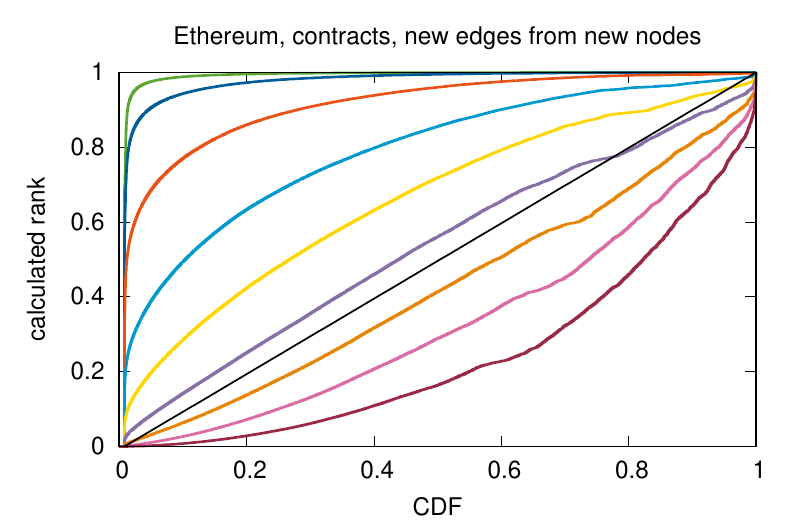} \\
	\includegraphics{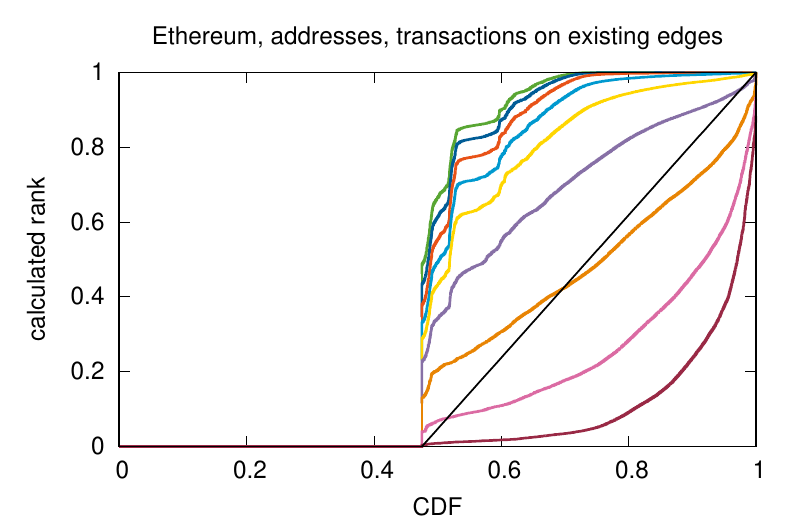}
	\includegraphics{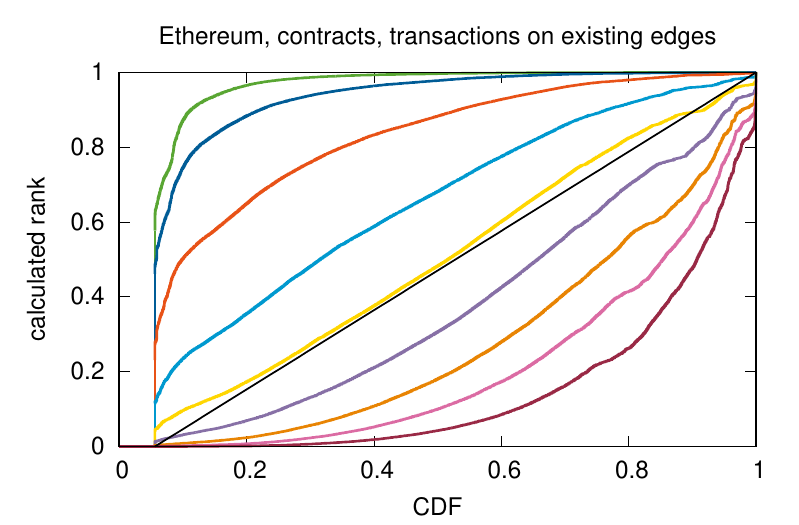} \\
	\includegraphics{figs_bal_legend}
	\caption{Testing for preferential attachment in Ethereum, assuming a 30 day ``lifetime'' for edges. The left column shows edges where the target is a regular address, while the right column shows edges where the target is a smart contract. Black lines show the expected ideal (i.e.~uniform) distribution. Kolmogorov-Smirnov differences from these distributions are shown in Fig.~\ref{ethereum_30d_exp}.}
	\label{ethereum_30d}
\end{figure*}

\begin{figure*}
	\centering
	\large Ethereum, one day edge lifetime \\
	\includegraphics{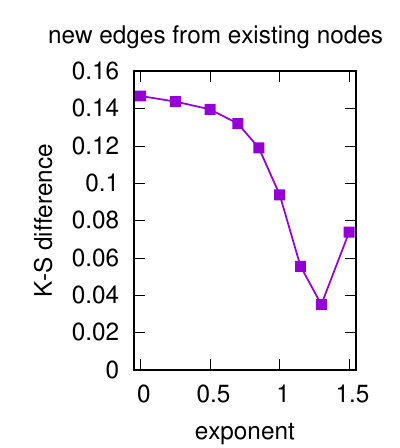}
	\includegraphics{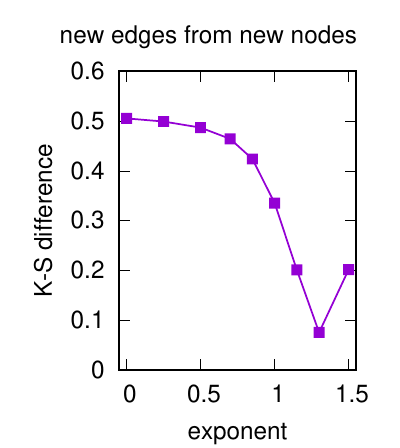}
	\includegraphics{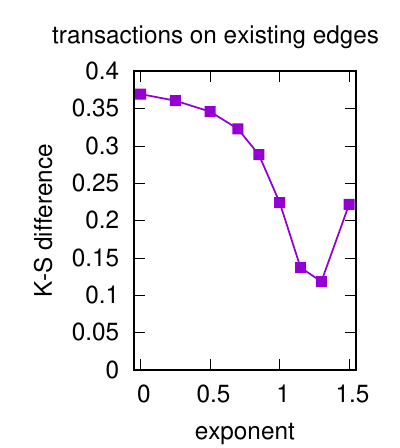}\\
	\includegraphics{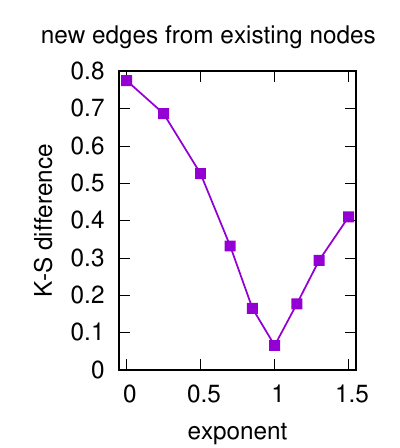}
	\includegraphics{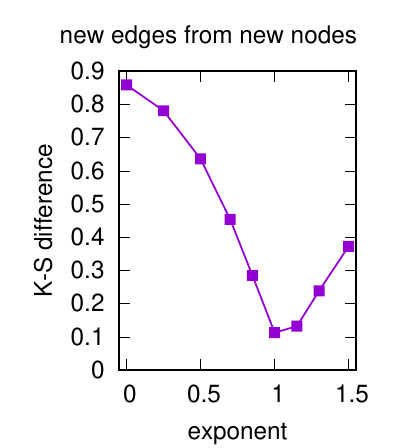}
	\includegraphics{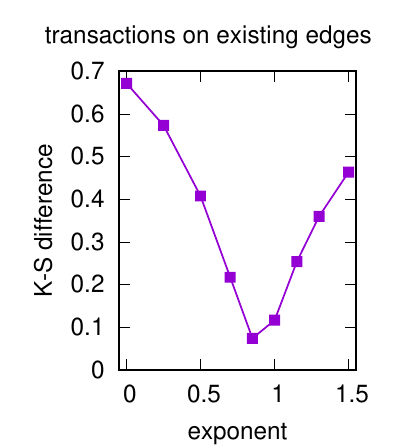}
	\caption{Kolmogorov-Smirnov differences from the presumed uniform distribution for the case of preferential attachment in Ethereum, assuming one day ``lifetime'' of edges, i.e.~for results displayed in Fig.~\ref{ethereum_1d}. Top row: results for transactions targeting addresses; bottom row: results for transactions targeting contracts.}
	\label{ethereum_1d_exp}
\end{figure*}

\begin{figure*}
	\centering
	\large Ethereum, 30 day edge lifetime \\
	\includegraphics{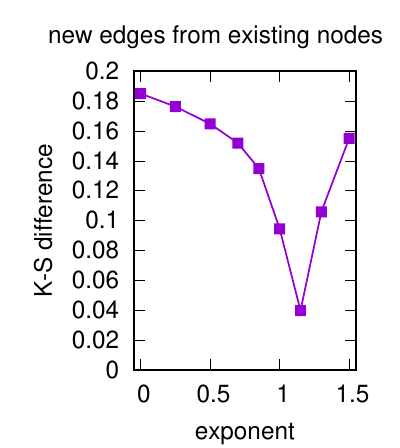}
	\includegraphics{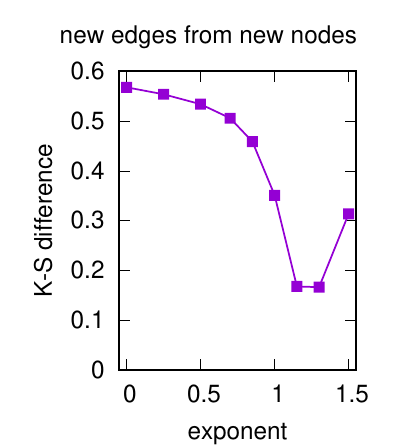}
	\includegraphics{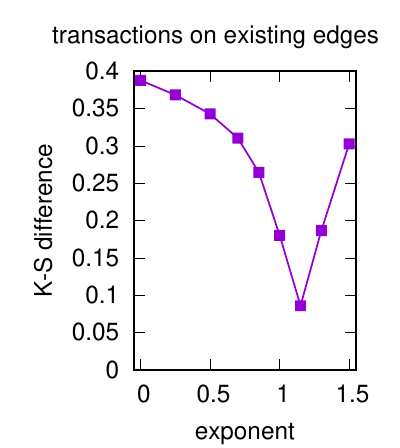}\\
	\includegraphics{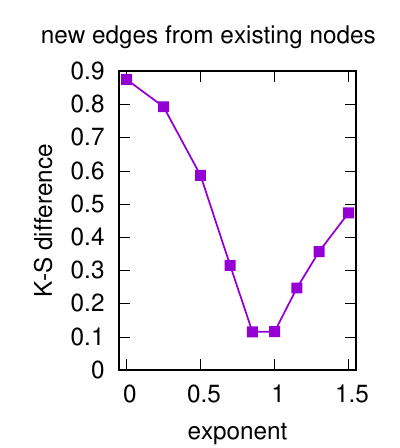}
	\includegraphics{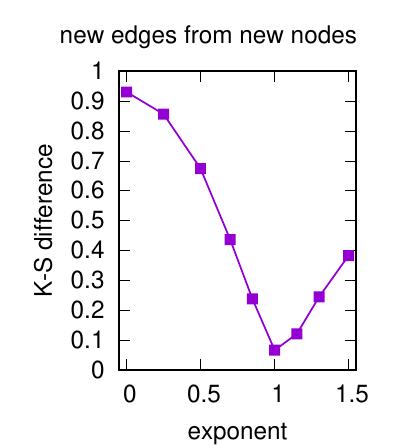}
	\includegraphics{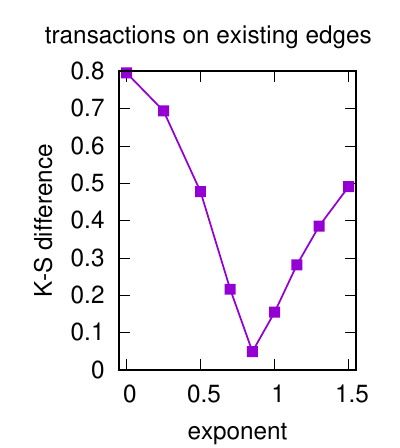}
	\caption{Kolmogorov-Smirnov differences from the presumed uniform distribution for the case of preferential attachment in Ethereum, assuming a 30 day ``lifetime'' of edges, i.e.~for results displayed in Fig.~\ref{ethereum_30d}. Top row: results for transactions targeting addresses; bottom row: results for transactions targeting contracts.}
	\label{ethereum_30d_exp}
\end{figure*}

Results are shown as the distribution of transformed ranks in Fig.~\ref{bitcoin2} for Bitcoin and in Figs.~\ref{ethereum_1d} and~\ref{ethereum_30d} for Ethereum, while we show the Kolmogorov-Smirnov distances from uniform distributions as a function of the $\alpha$ exponent in Figs.~\ref{bitcoin_exp_1d} and~\ref{bitcoin_exp_30d} for Bitcoin and in Figs.~\ref{ethereum_1d_exp} and~\ref{ethereum_30d_exp} for Ethereum.

\section{Efficient calculation of transformed rank values}

To evaluate statistics of preferential attachment, we need to calculate the following transformed rank for each edge linking event (either new edge creation or repeated transactions on the same edge):

\begin{equation}
	R \equiv \frac{\sum_{k = 0}^{k_\textrm{target}} n(k) k^\alpha}{\sum_{k = 0}^{k_\textrm{max}} n(k) k^\alpha}
	\label{eq:R}
\end{equation}

For a network with $N$ nodes, a naive implementation will have a runtime complexity of $O(N)$. If we have a total of $M$ events (with $M \sim 620$ million for Ethereum and $M \sim 4.5$ billion for Bitcoin), the total runtime complexity of evaluating the distribution of transformed rank values is $O(N M)$, assuming that updating node degrees is done in $O(1)$ time. Since performing this computation would be extremely slow even on modern hardware, we created an implementation that based on an augmented red-black tree that has a total runtime complexity of $O(M \log N)$.

Specifically, our implementation can be considered a generalized order-statistic tree. An order statistic tree is a binary search tree that allows calculating the rank of any element and finding an element with a given rank efficiently (in $O(\log N)$ time; for a formal introduction to binary search trees, see e.g.~\cite{algorithms}). A generalization of order statistic trees that allows the efficient calculation of the partial sum of \emph{any} value associated with its elements can be obtained in a straightforward way, by storing such partial sums as additional data in a suitably augmented binary search tree~\cite{Austern}. A more complete treatment on augmented binary search trees, and their usage for calculating transformed ranks was given as Appendix~A1 in Ref.~\cite{thesis}; in the following, we provide a summary of key concepts.

For our particular use case, we need to calculate the sums of the $k^\alpha$ values. In practice, we start with an implementation of a standard red-black tree~\cite{algorithms} that allows insertion and removal of nodes in $O(\log N)$ time complexity. We use the network degrees ($k$) as keys, and store $n(k)$, i.e.~the number of nodes with degree $k$ as mapped value in each node of the tree. Furthermore, we also store the partial sum of $n(k) k^\alpha$ in each node that corresponds to the subtree of that node. We ensure that these sums are recursively updated on each operation of the tree (this can be achieved in $O(\log N)$ time since each update needs to be only propagated upward until it reaches the root of the tree.

When a degree of a node in our transaction network changes, we find a node with the corresponding value in our red-black tree, decrease the stored count (or remove the tree node if it would reach zero), and recursively updated the stored partial sums. After this, we either add a new tree node with the new degree or increase the count if such a node already exists. Again, we take care to update partial sums.

When we need to calculate a transformed rank value for a target degree $k^*$, we first find a tree node with key $k^*$. Then we recursively calculate the sums of stored partial sums in the left subtrees of all nodes starting from the selected one up to the tree root node, according to Algorithm~\ref{alg:tree_sum}. This will give us the nominator in Eq.~(\ref{eq:R}), while the denominator is simply given by the partial sum value stored in the tree root node. Again, this operation can be carried out in $O(\log N)$ time complexity, since each level of the tree is visited only maximum once.

A general implementation of the augmented red-black tree used in the current work is available as Ref.~\cite{orbtree}; a somewhat specialized version along with code and scripts calculating transformed rank statistics is available as Ref.~\cite{patest_new}.

\begin{algorithm}
	\begin{algorithmic}[1]
		\State T: red-black tree storing the degree distribution and partial sums:
		\State \quad $x$: node of the tree corresponding to degree $k_x$, storing:
		\State \quad \quad $n_x$, the number of such network nodes
		\State \quad \quad $S_x$, the partial sum; $S_x = S_y + S_z + k_x^\alpha$ if $y$ and $z$ are children of $x$
		\State $k$: target degree of a transaction
		\State $x = $ T.find($k$) \Comment find the tree node with the given degree
		\State $R = 0$
		\While{True} \Comment Calculate the nominator of Eq.~(\ref{eq:R})
			\State $y = x$.left()
			\If{$y \neq $ T.nil()}
				\State $R = R + S_y$
			\EndIf
			\If{$x = $ T.root()}
				\State Break
			\EndIf
			\State $x = x$.parent()
		\EndWhile
		\State $x = $ T.root() \Comment Retrieve the denominator
		\State $R = R / S_x$
		\State Results: $R$, the transformed rank corresponding to a transaction targeting a node with degree $k$.
	\end{algorithmic}
	\caption{Algorithm calculating the transformed rank for one linking event.}
	\label{alg:tree_sum}
\end{algorithm}

\end{document}